\documentclass[manuscript]{emulateapj}

\usepackage{graphicx}

\lefthead{Jerjen et al.} 
\righthead{Main Sequence Star Populations in the Virgo Overdensity Region} 
\begin{document}      

\title{Main Sequence Star Populations in the Virgo Overdensity Region\footnote{Based on data collected 
at Subaru Telescope, which is operated by the National Astronomical Observatory of Japan.}}

\author{
H.~Jerjen\altaffilmark{1}, 
G.S.~Da Costa\altaffilmark{1}, 
B.~Willman\altaffilmark{2},
P.~Tisserand\altaffilmark{1}, 
N.~Arimoto\altaffilmark{3,4},
S.~Okamoto\altaffilmark{5},
M.~Mateo\altaffilmark{6},
I.~Saviane\altaffilmark{7},
S.~Walsh\altaffilmark{8},
M.~Geha\altaffilmark{9}, 
A.~Jord\'an\altaffilmark{10,11}, 
E.~Olszewski\altaffilmark{12},
M.~Walker\altaffilmark{13},
M. Zoccali\altaffilmark{10,11},
P.~Kroupa\altaffilmark{14}
}      
\affil{$^1$Research School of Astronomy \& Astrophysics, The Australian National University, Mt Stromlo Observatory, 
via Cotter Rd, Weston, ACT 2611, Australia}
 \affil{$^2$Haverford College, Department of Astronomy, 370 Lancaster Avenue, Haverford, PA 19041, USA}
 \affil{$^3$National Astronomical Observatory of Japan, Subaru Telescope, 650 North A'ohoku Place, Hilo, 96720 USA}
  \affil{$^4$The Graduate University for Advanced Studies, Department of Astronomical Sciences, Osawa 2-21-1, Mitaka, Tokyo, Japan}
 \affil{$^5$Kavli Institute for Astronomy and Astrophysics, Peking University, Beijing 100871, China}
 \affil{$^6$Department of Astronomy, University of Michigan, Ann Arbor, MI, USA}
 \affil{$^{7}$European Southern Observatory, Casilla 19001, Santiago 19, Chile}
\affil{$^8$Australian Astronomical Observatory, PO Box 915, North Ryde, NSW 1670, Australia}
\affil{$^9$Astronomy Department, Yale University, New Haven, CT 06520, USA}
\affil{$^{10}$Departamento de Astronom\'ia y Astrof\'isica, Pontificia Universidad Cat\'{o}lica de Chile, 7820436 Macul, Santiago, Chile}
\affil{$^{11}$The Milky Way Millennium Nucleus, Av. Vicu\~{n}a Mackenna 4860, 782-0436 Macul, Santiago, Chile}
\affil{$^{12}$Steward Observatory, The University of Arizona, Tucson, AZ, USA}
\affil{$^{13}$Harvard-Smithsonian Center for Astrophysics, 60 Garden Street, Cambridge, MA 02138, USA}
 \affil{$^{14}$Argelander Institute for Astronomy, University of Bonn, Auf dem H\"ugel 71,D-53121 Bonn, Germany}

\begin{abstract} 
We present deep color-magnitude diagrams for two Subaru Suprime-Cam fields in the Virgo 
Stellar Stream(VSS)/Virgo Overdensity(VOD) and compare them to a field centred on the 
highest concentration of Sagittarius (Sgr) Tidal Stream stars in the leading arm, Branch\,A of the bifurcation. 
A prominent population of main sequence stars is detected in all three fields and can be traced as faint 
as $g\approx24$\,mag. Using theoretical isochrone fitting we derive an age of $9.1^{+1.0}_{-1.1}$\,Gyr, a median 
abundance of [Fe/H]=$-0.70^{+0.15}_{-0.20}$\,dex and a heliocentric distance of $30.9\pm3.0$\,kpc for the main 
sequence of the Sgr Stream Branch\,A. The dominant main sequence populations in the two VSS/VOD fields ($\Lambda_\sun\approx265^\circ$, 
$B_\sun\approx13^\circ$) are located at a mean distance of $23.3\pm1.6$\,kpc and have an age $\sim 8.2$\,Gyr
and an abundance [Fe/H]=$-0.67^{+0.16}_{-0.12}$\,dex similar to the Sgr Stream stars. These statistically 
robust parameters, derived from photometry of 260 main sequence stars, are also in good agreement with the age 
of the main population in the Sagittarius dwarf galaxy ($8.0\pm1.5$\,Gyr). They also agree with the peak in the metallicity 
distribution of $2-3$\,Gyr old M-giants, [Fe/H]$\approx -0.6$\,dex, in the Sgr leading arm north. 
We then compare the results from the VSS/VOD fields with the Sgr Tidal Stream model by Law \& Majewski based on 
a triaxial Galactic halo shape that is empirically calibrated with SDSS Sgr A-branch and 2MASS M\,giant stars. 
We find that the most prominent feature in the CMDs, the main sequence population at 23\,kpc, is not explained 
by the model.  Instead the model predicts in these directions a low density filamentary structure of Sgr debris stars 
at $\sim 9$\,kpc and a slightly higher concentration of Sgr stars spread over a heliocentric distance range of $42-53$\,kpc.  
At best there is only marginal evidence for the presence of these populations in our data.  Our findings then suggest 
that while there are probably some Sgr debris stars present, the dominant stellar population in the Virgo Overdensity 
originates from a different halo structure that has almost identical age and metallicity as some sections of the Sgr tidal stream.
\end{abstract}         
\section{Introduction}  
The first decade of wide-field digital imaging has revolutionised the way the
3-D structure of the Milky Way can be mapped. The tomographic studies 
enabled by programs such as the Sloan Digital Sky Survey \citep[SDSS,][]{York00} or 
the Two Micron All Sky Survey \citep[2MASS,][]{Skrutskie06} 
have not only uncovered members from a new family of ultra-faint satellite galaxies but also revealed 
strong evidence for the presence of a significant amount of substructure in the halo 
of the Milky Way in the form of large scale stellar streams. 
Understanding the nature and origin of these satellite galaxies and their tidal debris
has major implications for the validity of cosmological models \citep[e.g.][]{Kroupa10}. 
The most striking 
satellite galaxy currently undergoing tidal disruption is the 
Sagittarius dwarf \citep{Ibata94, Ibata01a, MartinezDelgado01, Majewski03}. This galaxy is 
located approximately 16\,kpc from the Galactic center \citep{Kunder09}, but the debris from the interaction 
with the Milky Way has been traced from 16--90\,kpc galactocentric distance across large areas of the sky 
\citep{Majewski03, Newberg03, Belokurov06b, Correnti10}.
\cite{Layden00} found evidence for multiple epochs of star formation in the Sgr dwarf 
with the principal star formation epochs at 11, 5, and 0.5--3 Gyr and associated
mean abundance values of [Fe/H]=$-1.3$,$- 0.7$, and $-0.4$, respectively. This picture 
was subsequently refined by \cite{Bellazzini06} finding that more than 80\,precent of 
Sgr stars belong to a relatively metal-rich ([Fe/H]$\sim-$0.7\,dex) and intermediate-old age ($8.0\pm1.5$\,Gyr)
population called Pop\,A, after \cite{Bellazzini99a} and \cite{Monaco02}.

Apart from major tidal streams, the analysis of large samples of tracers of halo 
substructure have further revealed the presence of numerous smaller stellar overdensities 
possibly associated with unknown fainter streams or merger events in the halo. In particular 
for the region of the Virgo constellation several detections of stellar overdensities have been 
reported covering the distance range 4--20\,kpc.  
Most noticeable is a significant concentration of 
RR Lyrae (RRL) stars at $\alpha = 12.4$\,h \citep{Vivas01, Vivas02, Vivas03, Zinn04, Ivezic05} located at $\approx$19\,kpc 
from the Sun \citep{Duffau06, Newberg07, Prior09a}. This clump was 
also detected as an excess of metal-poor, old F-type main sequence turnoff stars by 
\cite{Newberg02} and as a distinct stellar clustering labeled Vir\,Z by \cite{Walsh09}. 
\cite{Duffau06} showed via spectroscopy that the majority of the RRL and 
blue horizontal branch stars in the overdense region define a kinematically cold
($\sigma$ $\approx$ 20 kms$^{-1}$)  feature
centered on a Galactic rest frame velocity V$_{GSR}$ of $\sim$100 kms$^{-1}$, which 
they labeled the ``Virgo Stellar Stream" (VSS). 

Furthermore, \cite{Juric08} and more recently \cite{Bonaca12} detected a density enhancement over 2000 sq degrees of sky toward the Virgo 
constellation by means of photometric parallax distance estimates of SDSS stars. 
They named this large scale feature the ``Virgo Overdensity" (VOD). The VOD is estimated 
to have a distance of $6-20$\,kpc \citep{Juric08, Vivas08, Keller09}.
In the following, we will use the term ``VSS" when we are referring to the feature identified kinematically and 
abundance-wise, and use the term ``VOD" for the spatial overdensity.

\cite{MartinezDelgado04, MartinezDelgado07} speculated
that the Virgo stellar overdensity might be related to the complex debris structure of the trailing arm of the 
Sagittarius stream, a scenario that is supported by the Sgr Stream models for this region of the sky 
\citep{Law05, Fellhauer06} assuming the Galactic halo potential has a spherical or 
oblate shape. Subsequent spectroscopic 
follow-up studies of a sample of these RRLs 
by \cite{Prior09a} not only found further RRLs associated with the VSS, both
in terms of velocity and abundance, but also
revealed a population of metal-poor RRLs with large negative V$_{GSR}$ velocities.  These stars
were taken to indicate the likely presence of a population of Sgr leading tidal tail stars, which are 
expected to have such velocities in this region.  \cite{Prior09b} also suggested that Sgr trailing 
debris may make a contribution to the population at positive V$_{GSR}$ velocities, though it
was unlikely to fully account for the VSS feature.  \cite{Chou07, Chou10} have argued from a
chemical abundance point-of-view for the presence of Sgr debris stars in this region.   

The aim of the present paper is to provide a robust estimate of the mean age of the stellar 
population that dominates the VSS/VOD region by means of deep imaging in two directions 
close to the 12.4h clump. For that purpose we observed two Subaru Suprime-Cam 
fields covering 0.25 square degrees each, centered around $\alpha_{2000}=12^h20^m18^s$,  
$\delta_{2000}=-01^d21^m00^s$ and $\alpha_{2000}=12^h47^m58^s$,
$\delta_{2000}=-00^d45^m00^s$, respectively. We also observed 
a field $19^\circ$ away, in the direction of the highest star density of the Sagittarius Branch\,A leading 
arm. All three fields were identified by \cite{Walsh09} as having statistically significant 
overdensities of point sources. 

In Section\,2 we describe the data acquisition, reduction and photometric calibration. 
The analysis of the CMDs is presented in Section\,3. In Section\,4 arguments are compiled 
for the interpretation of our measurements in the context of the VOD being dominated by the Sgr Tidal Stream. 
An alternative interpretation of the CMDs naturally arises from the comparison with the predictions 
of the Sagittarius dwarf--Milky Way halo interaction model. This is tested in Section\,5. Finally, in Section\,6
we summarize and discuss our conclusions.

\begin{table}\centering\label{field_coords}
\caption{Coordinates of the Sgr Tidal Stream and the two Virgo fields}
\begin{tabular}{lcccccc}
          & $\alpha_{2000}$ & $\delta_{2000}$  & $l$ & $b$  & $\Lambda_\sun$ & $B_\sun$ \\
Field & (deg) & (deg)  & (deg) &  (deg) & (deg) & (deg)\\
\hline
1145$+$13 &   176.288 & +13.95    & 250.07 & +69.68  & 247.8 & 5.2 \\  
1220$-$01& 185.077 & $-$01.35 &286.95 & +60.55  & 262.1 & 15.3 \\  
1247$-$00 & 191.992 & $-$00.75  &301.07 & +62.11 & 268.6 & 11.4 \\ 
\hline\\
\end{tabular}
\end{table}

\section{Data Acquisition, Photometry and Calibration} 
We obtained deep $g$, $r$ CCD images of the three fields (Table\,\ref{field_coords}) using the Suprime-Cam 
\citep{Miyazaki02} on the Subaru Telescope during nights of 2009 February 22 and 23 
(PI. N. Arimoto).  The Suprime-Cam consists of a 5 $\times$ 2 array of 2048 $\times$ 4096 CCD detectors 
and provides a field-of-view of 34\arcmin $\times$ 27\arcmin with a pixel scale of 0.202\arcsec. The first night 
was photometric and the second night partially clear with the median seeing during both nights at 0.8\arcsec. 
Each field was observed in a  dithered series of 5 $\times$ 280\,s exposures in the SDSS $g$ band and 
10 $\times$ 200\,s in $r$. Data were processed using the pipeline software SDFRED dedicated to the Suprime-Cam \citep{Yagi02, 
Ouchi04}.  Each image was bias-subtracted and trimmed, flat-fielded, distortion and atmospheric 
dispersion corrected, sky-subtracted, and combined in the usual manner. The astrometric calibration of each passband 
was based on a general zenithal polynomial projection derived from astrometric standard stars selected from online 
USNO catalog\footnote{http://ftp.nofs.navy.mil/data/fchpix/}.  

\begin{figure}[t]\centering
\includegraphics[width=4.9cm]{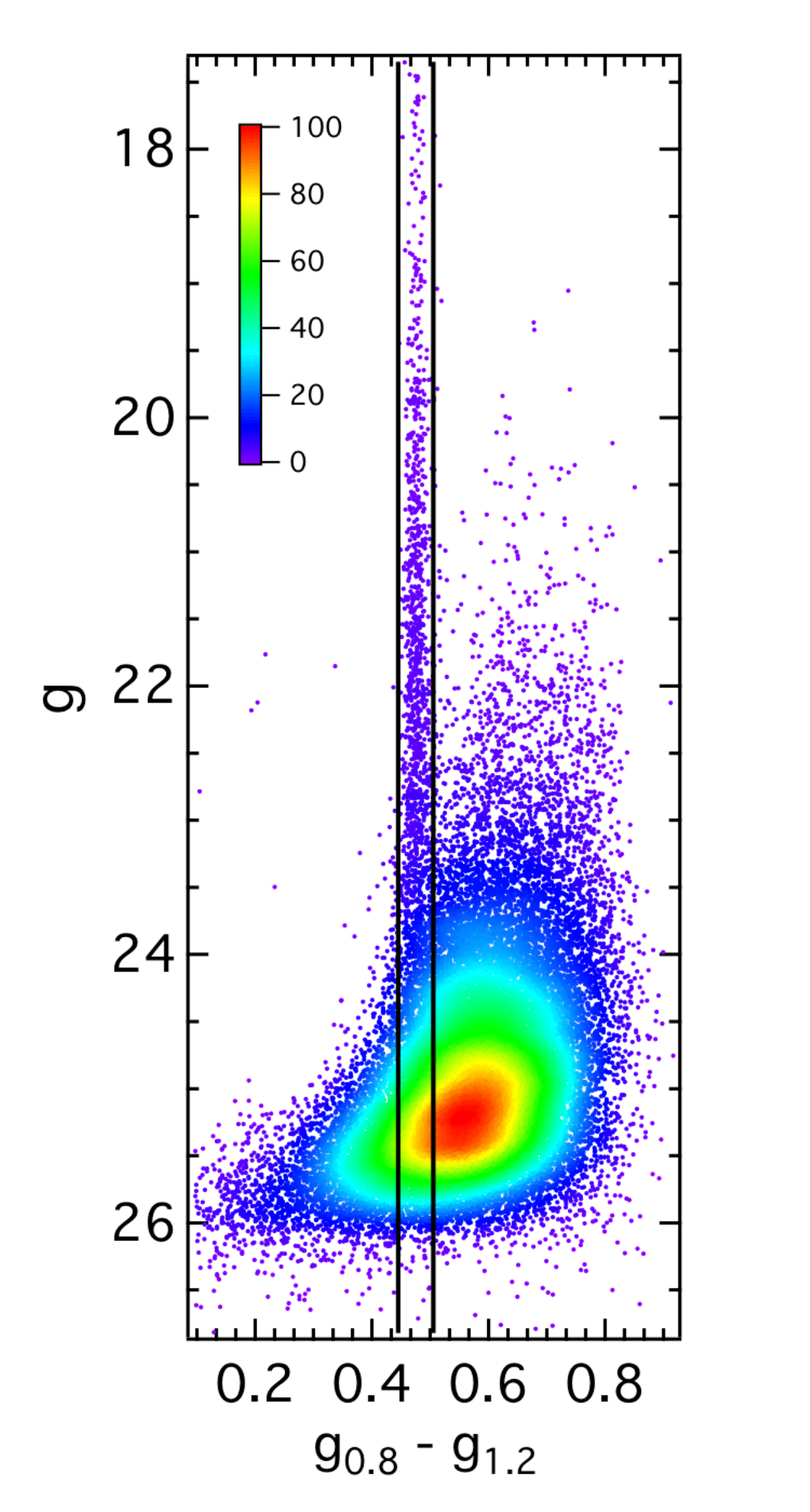}\hspace*{-0.8cm}
\includegraphics[width=5cm]{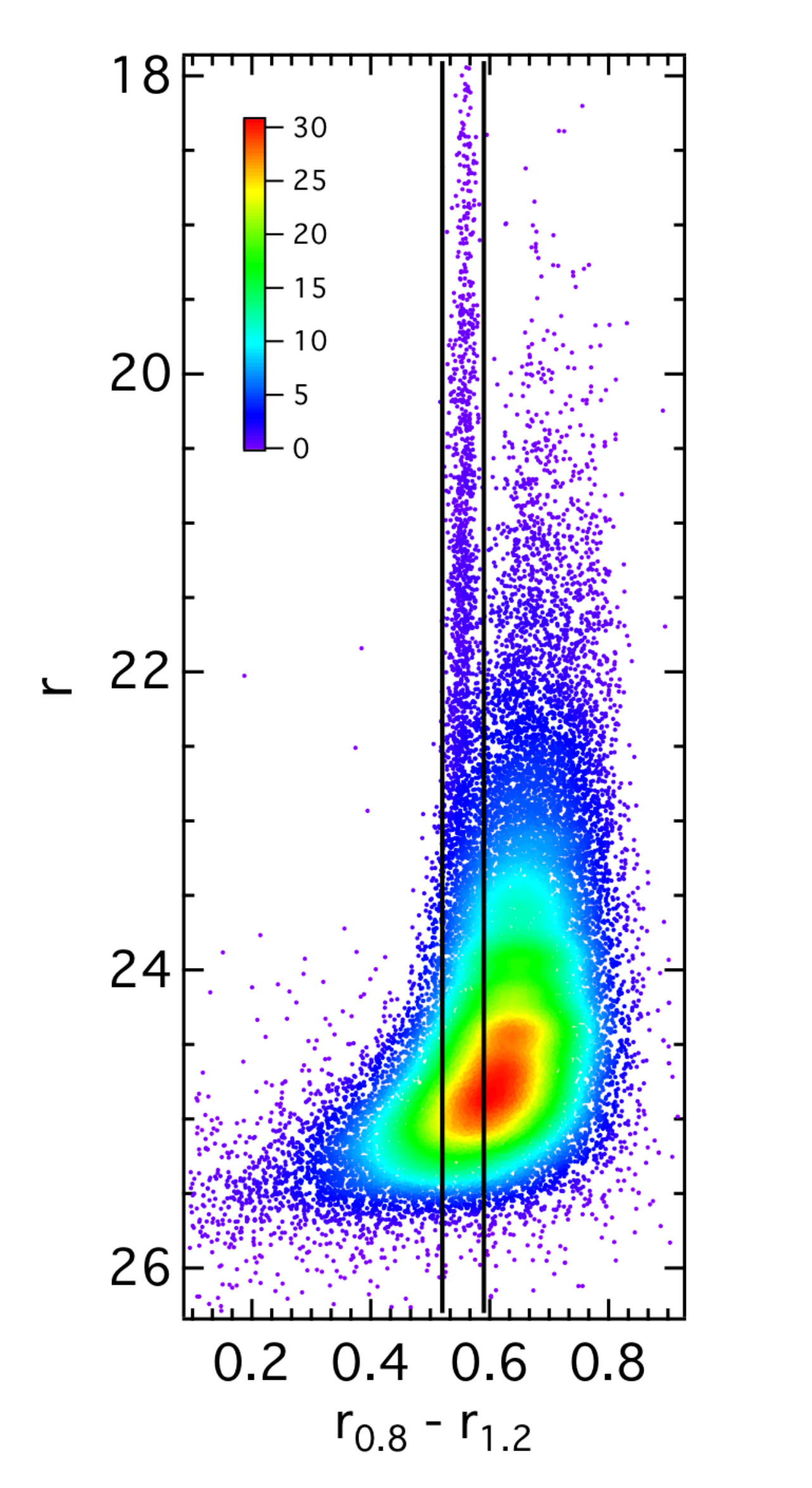}
\caption{An example of star-galaxy separation: the aperture flux ratio versus total flux for all 
detected sources in the  $1220-01$ field. Unsaturated stars populate well-defined vertical 
bands  $0.45<g_{0.8}-g_{1.2}<0.51$ and $0.52<r_{0.8}-r_{1.2}<0.59$, respectively. 
Selecting objects in these two narrow regions enables a fair star-galaxy separation 
down to $g\sim$23.7\,mag. Fainter objects can be stars or unresolved background galaxies 
(see Fig.\ref{rawcmds}). 
Sources are color-coded 
with the local number density.} 
\label{sgseparation}
\end{figure}

SourceExtractor \citep{Bertin96} was employed to identify and analyse sources 
down to $g\sim 26$\,mag. The source catalogs include PSF and flux measurements inside two 
circular apertures (diameter: 0.8\arcsec and 1.2\arcsec) optimized for the seeing. 
We have tied our photometry to the SDSS photometric system by matching all the point sources from
SDSS observed in our fields with our $g$-band and $r$-band detections.
The magnitude zero-points with respect to the Galactic-extinction corrected 
SDSS photometry were obtained by calculating a 3-$\sigma$ clipped mean of the magnitude 
differences of the unsaturated SDSS stars that were detected in our Subaru data. The 
comparison of typically 100--200 SDSS stars in the magnitude interval $17.6<g,r<22.5$ for each 
calibration made these zero-points statistically robust with uncertainties less than 0.01\,mag.

To separate stars from background galaxies and other non-stellar objects, we compare 
the fluxes in the two different apertures. The PSF geometry places stellar objects 
in a well-defined, narrow region in the magnitude--flux ratio plane. Such a plot is shown 
in Fig.~\ref{sgseparation} for the $1220-01$ field. Unsaturated stars populate vertical 
bands centered at $g_{0.8}-g_{1.2}=0.52$ and $r_{0.8}-r_{1.2}=0.54$, respectively. 
Selecting only objects with values in these two bands ensures a fair star-galaxy 
separation in our images down to $g\sim23.5-24.0$\,mag, a limit that depends on the seeing. 
At fainter magnitudes the signal-to-noise gets too low and unresolved background galaxies blend 
into the region leading to an increasing level of contamination. All color-magnitude diagrams 
(CMDs; $(g-r)_0$ vs. $g_0$) shown in our study are based on aperture photometry with a 0.8\arcsec 
diameter aperture calibrated to the SDSS system and are restricted to objects that fall 
simultaneously into both stellar flux ratio intervals as determined for each field individually.

Artificial star tests were conducted to estimate the detection completeness. 
A catalogue of uniformly distributed simulated stars was added to each 
image based on measured PSF parameters and noise properties. Photometry of all 
the stars was then performed and typical photometric errors derived as the 
difference between input and computed magnitude. We calculated the fraction 
of simulated stars recovered at different locations across the image to test whether 
stellar crowding is present and how it affects the completeness of the observations. 
The derived completeness curves (recovery rates as a function of magnitude) 
did not vary across the FoV and showed a steep gradient where the 
50\% source completeness in the photometry is approximately 0.5\,mag brighter than 
the limiting magnitude of $g_{lim}\sim 26$\,mag.

\begin{figure*}[thb]\centering  
\includegraphics[width=4.8cm]{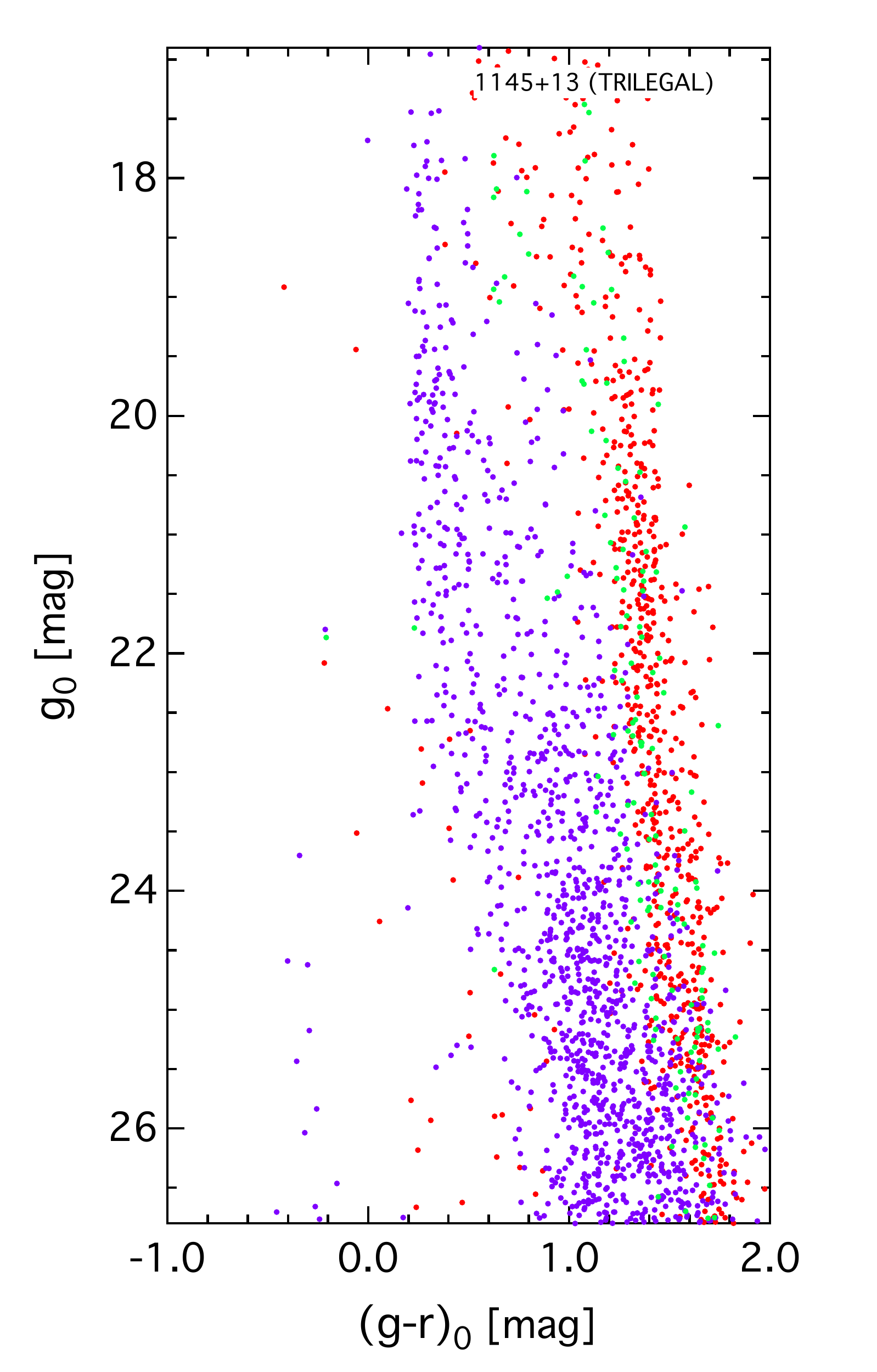}\hspace{-0.5cm}
\includegraphics[width=4.8cm]{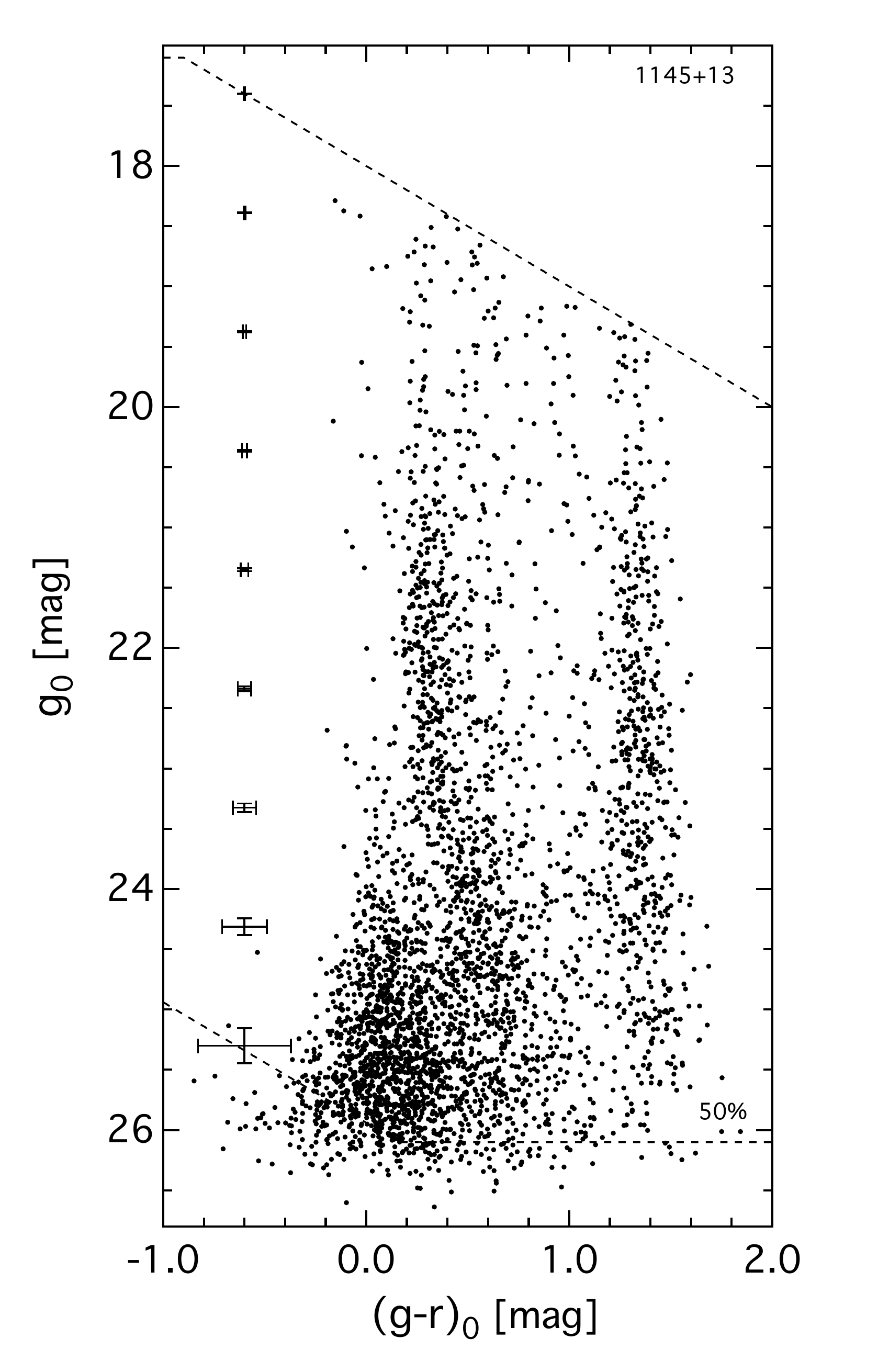}\hspace*{-0.5cm}
\includegraphics[width=4.8cm]{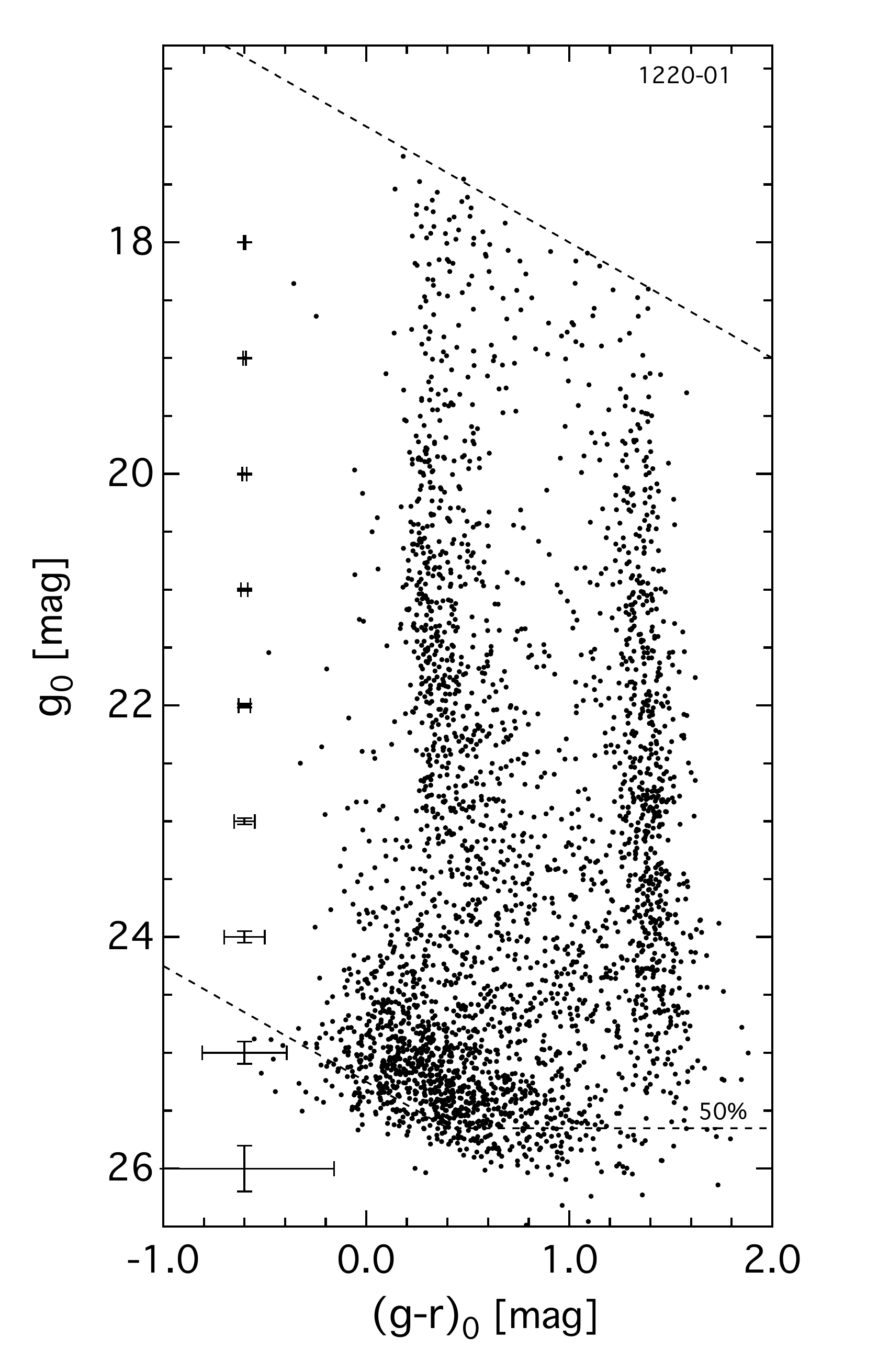}\hspace*{-0.5cm}
\includegraphics[width=4.8cm]{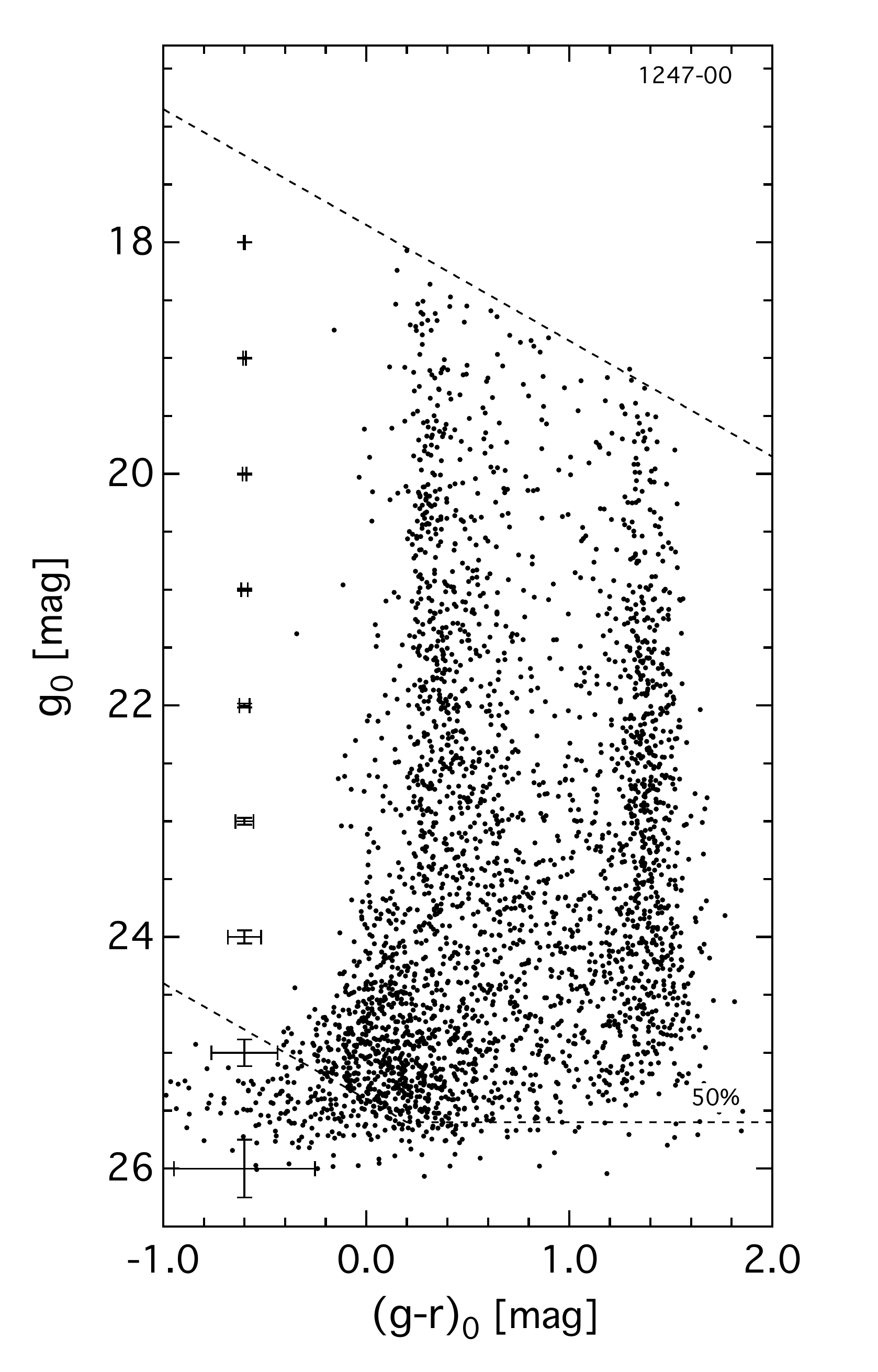}\hspace*{-0.5cm}
\caption{(a) Color-magnitude diagram of the Galactic field in the direction of 
1145+13 simulated with the TRILEGAL\,1.5 code. Different stellar populations are 
coded in color (thin disk - red; thick disk - green; halo - blue).
(b--d) Reddening corrected color-magnitude diagrams of the three Subaru fields with 
representative photometric errors plotted on the left side of the CMD.
(b) CMD of the field $1145+13$ in the direction of the Sagittarius Stream A branch. 
A prominent main sequence population is visible with the turn-off at $g\sim 21.0$\,mag. 
(c) CMDs of the field 1220$-$01 and (d) 1247$-$00, respectively.
Similar to the 1145+13 field a prominent main sequence population is visible with the turn-off at 
$g\sim 20.0-20.5$\,mag. Unresolved background galaxies appear as plume below $g=24$ 
in the color interval $-0.4<g<0.6$.\label{rawcmds}}
\end{figure*}

\section{Color-Magnitude Diagrams}
Over the observed FoV of 918 sq arcmin several thousand sources satisfy the aperture flux selection.
The three CMDs (Fig.~\ref{rawcmds}) show two distinct vertical plumes of Galactic stars 
reflecting the separation of the halo and thick disk stars at a color of $(g-r)_0\sim 0.3$ and the thin 
disk stars at $(g-r)_0\sim 1.35$ \citep[see Fig.\ref{rawcmds}a or ][]{Chen01}. Below 
$g_0=24$\,mag and around $-0.2<(g-r)_0<0.8$ an increasing number of unresolved background 
galaxies start dominating the CMDs. The Subaru photometry typically reaches down to a limiting 
magnitude of $g_{lim}\sim 26$\,mag but we will restrict our analysis to the interval $18<g<24$. 

The CMDs were statistically decontaminated from unwanted Galactic stars using model CMDs generated for 
each field with the TRILEGAL code \citep{Girardi05, Vanhollebeke09}. As an example, the simulated CMD of the Galactic field 
in the direction of 1145+13 is shown in Fig.~\ref{rawcmds}a. A comparison of the numbers of observed and predicted 
Galactic stars in the magnitude-color intervals $1.2<g-r<1.8$, $20<g<24$ finds good agreement (Table\,\ref{cmdcomp}):

\begin{table}[h]
\caption{Star counts in the box $1.2<g-r<1.8$, $20<g<24$}
\begin{center}
\begin{tabular}{ccc}
Field&  Observed  & TRILEGAL  \\\hline
1145$+$13  & 367 & 395 \\
1220$-$01  & 536 &510 \\
1247$-$00  & 546 & 568\\\hline\label{cmdcomp}
\end{tabular}
\end{center}
\end{table}

The cleaning procedure involves a process, 
in which a point source A in the CMD of 1145$+$13, 1220$-$00, or 1247$-$01
is removed if there is a source B in the Galactic CMD that lies within the 3$\sigma$ photometry uncertainty ellipse of 
A. Source B is also removed from the control field catalog for the rest of the process to avoid repeated use. 
If more than one source is found in the ellipse, the closest to A is discarded. The decontaminated CMDs are 
shown in Fig.~\ref{cmdfits}. We note that the stars in the simulated CMD represent a single realisation 
of the contamination, which explains why not all Galactic stars were removed. Cleaning efficiencies of 
68 percent (1145$+$13), 63 percent (1220$-$00), and 59 percent (1247$-$01) were achieved in the
``pure field" part of the CMD ($1<g-r<2$, $20<g<24$). For the $0<g-r<1$ color interval where the star
density is higher compared to a pure field the cleaning process is expected to be more efficient because of
the higher probability that an observed star matches with a star in the TRILEGAL CMD and thus 
is removed.

\subsection{Field $1145+13$ in the Sgr Stream Leading Arm}
\begin{figure*}\centering  
\includegraphics[width=5.7cm]{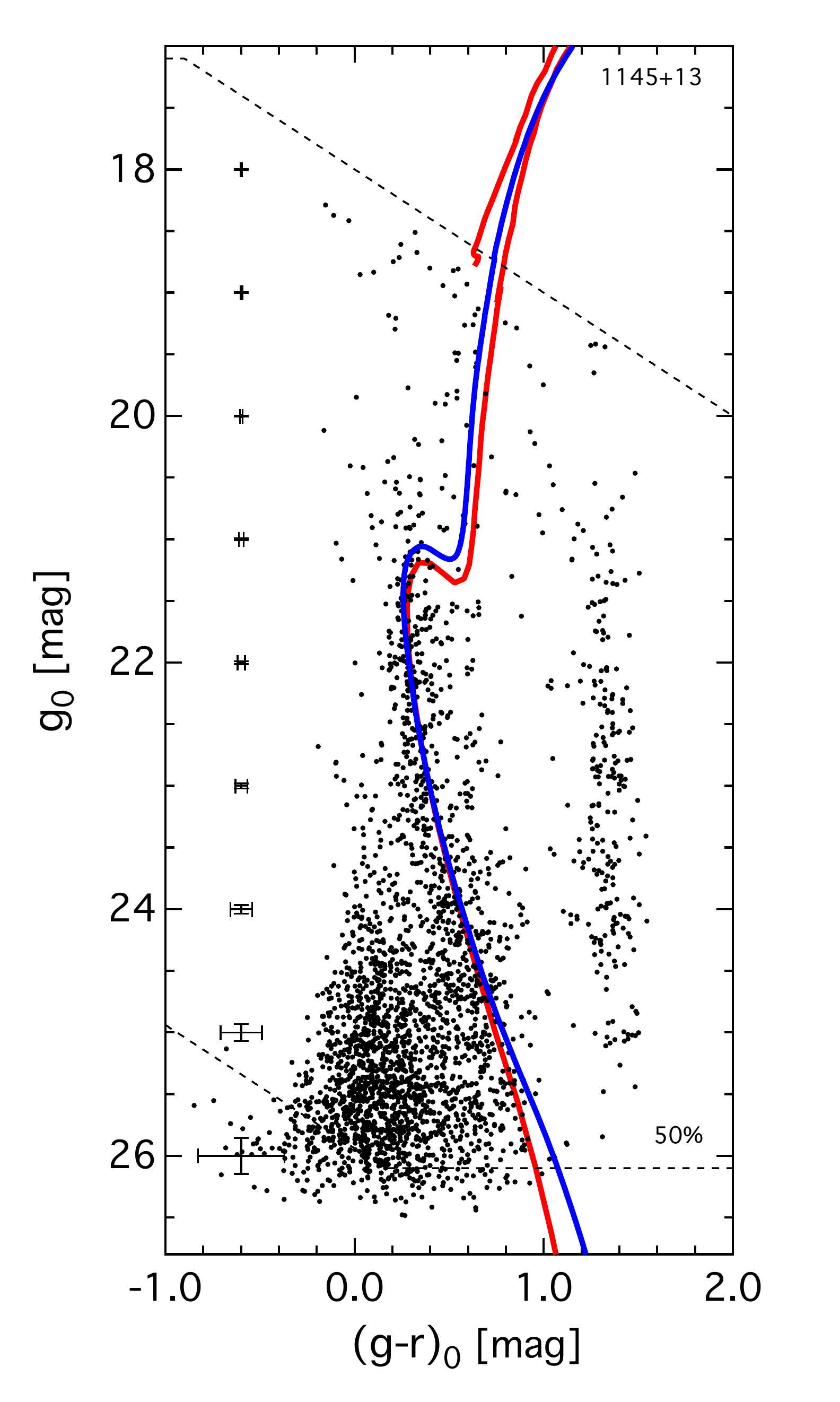}\hspace*{-0.5cm}
\includegraphics[width=5.7cm]{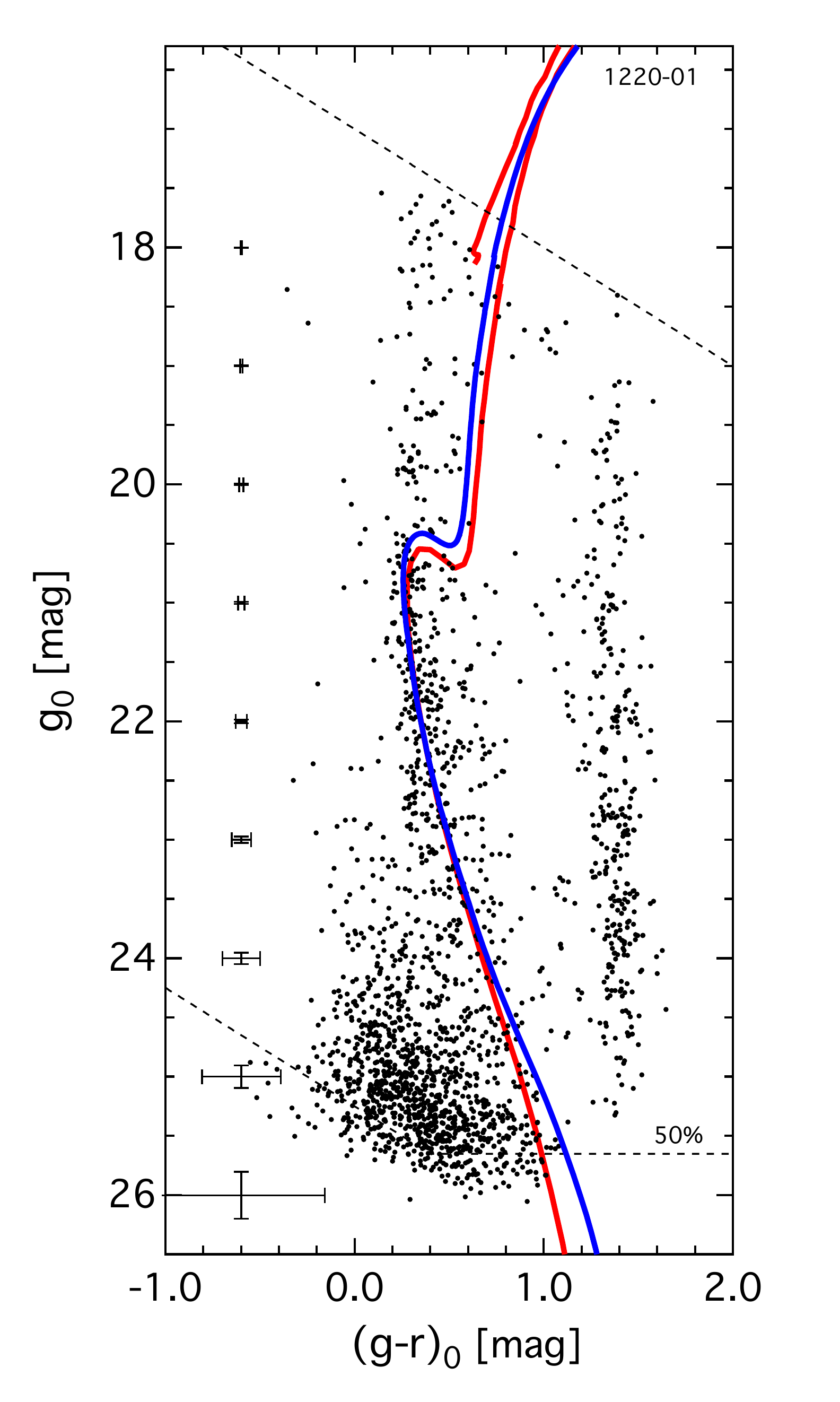}\hspace*{-0.5cm}
\includegraphics[width=5.7cm]{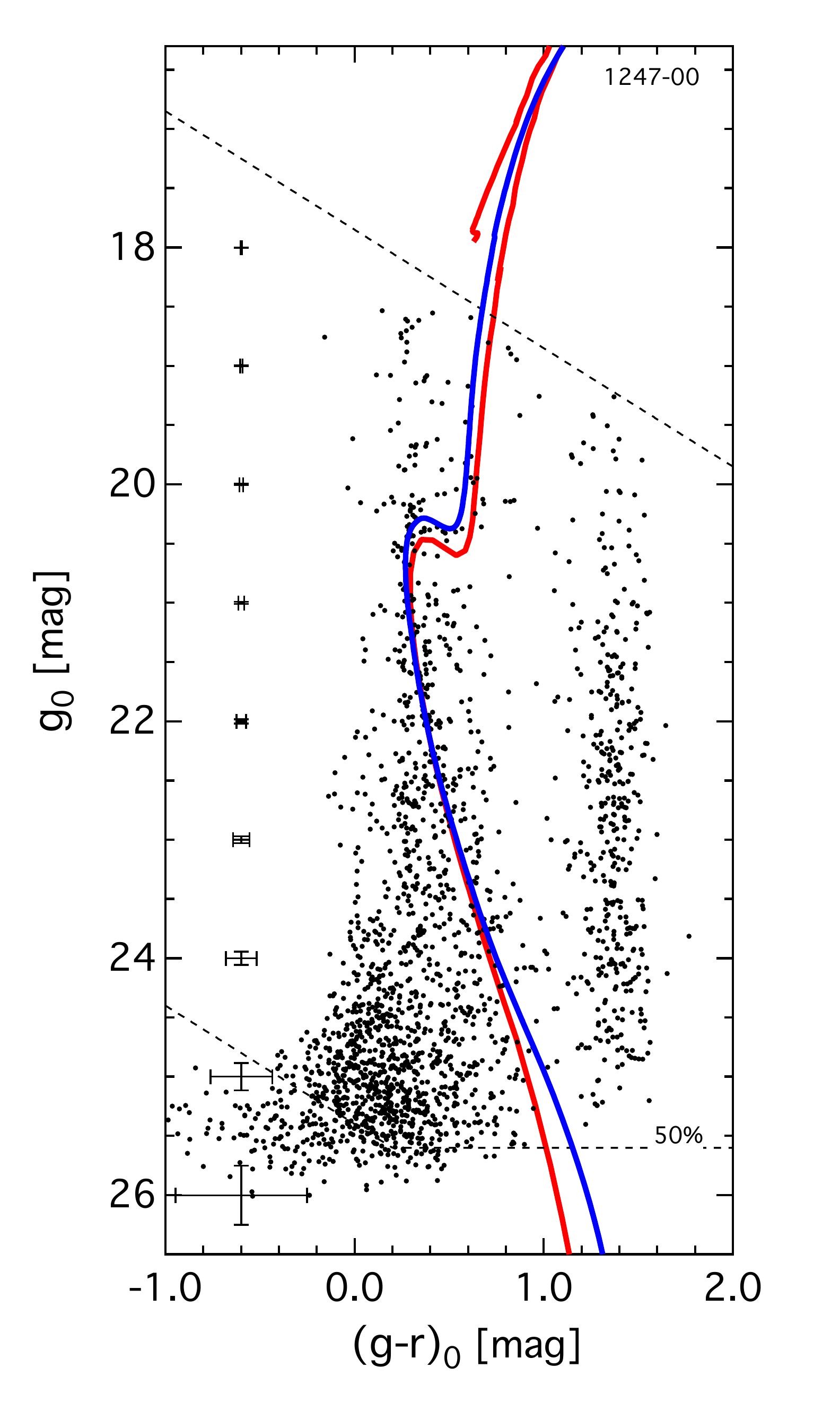}\hspace*{-0.5cm}
\caption{Foreground subtracted CMDs with the best-fitting Dartmouth isochrones (blue line) 
to the main sequence population 
(a) 1145+13: has a metallicity of [Fe/H]=$-0.70$\,dex and an age of 9.1\,Gyr. 
The associated heliocentric distance (modulus) is 30.9\,kpc ($m-M=17.45$\,mag). 
(b) 1220$-$01: has a metallicity of [Fe/H]=$-0.66$\,dex and an age of 7.9\,Gyr. 
The associated heliocentric distance (modulus) is 24.3\,kpc ($m-M=16.93$\,mag). 
(c) 1247$-$00: has a metallicity of [Fe/H]=$-0.68$\,dex and an age of 8.5\,Gyr. 
The associated heliocentric distance (modulus) is 22.2\,kpc ($m-M=16.73$\,mag). 
The Padova isochrones with the same ages, metallicities, and distances
are shown in red for comparison.\label{cmdfits}}
\end{figure*}

The tidal stream of the Sgr dwarf galaxy in the SDSS data visibly diverges westwards of $\alpha\approx 190^\circ$ to give a prominent 
bifurcation known as Branches\,A and B  \citep{Belokurov06b}. Our field $1145+13$ is located at equatorial 
coordinates $\alpha_{2000}=176.29^\circ$, $\delta_{2000}=13.95^\circ$ in the direction 
of the highest star density of the Sagittarius Branch\,A 
as defined by \cite{Belokurov06b}. In line with expectations, our CMD (Fig.~\ref{rawcmds}b) exhibits 
a diagonal ridge of main sequence stars in the interval $21<g<25.5$, a feature that is even more pronounced in the 
statistically decontaminated version of the CMD (Fig.~\ref{cmdfits}). The Sgr Stream population has a measured 
MSTO color of $(g-r)_0\sim0.25$. We derive the age, [Fe/H], and distance of 
the Sagittarius Stream using a maximum likelihood method which closely follows \citet{frayn02} 
and was employed to analyse the CMD of Segue\,3 \citep{Fadely11}. The fundamental assumption is that 
the data is dominated by a single age/metallicity population.
For the procedure, a suite of isochrones are fitted to a sample of stars, assigning to each a bivariate Gaussian probability function whose variance 
is set by the associated photometric errors $\sigma_g$ and $\sigma_{(g-r)}$.  We apply this analysis on all stars in our decontaminated CMD.
For a given isochrone $i$ we compute the likelihood 

\begin{eqnarray}
\mathcal{L}_i=\prod_j p(\{g,g-r\}_j|i,\{g,g-r\}_{ij},{\rm DM}_{i}) ,
\end{eqnarray}

\noindent where $p$ is defined as: 

\footnotesize

\begin{eqnarray}
\label{eqn:eachlike}
&&p(\{g,g-r\}_j|i,\{g,g-r\}_{ij},{\rm DM}_{i})=\frac{1}{2\pi\sigma_{g_j}\sigma_{(g-r)_j}} \times  \\ 
&&\exp\left(-\frac{1}{2}\left[\left(\frac{g_j-(g_{ij}+{\rm DM}_{i})}{\sigma_{g_j}}\right)^2 + \left(\frac{(g-r)_j-(g-r)_{ij}}{\sigma_{(g-r)_j}}\right)^2 \right]\right)\nonumber .
\end{eqnarray}

\normalsize

\noindent For each star $j$, $\{g,g-r\}_{ij}, {\rm DM}_{i}$ are the magnitude, color, and de-reddened distance 
modulus values for isochrone $i$ that maximize the likelihood of the entire data set $\{g,g-r\}_{j}$ in Equation 
\ref{eqn:eachlike}.  We take an approximate solution to finding the values of $\{g,g-r\}_{ij}$ and $\rm DM_{i}$ by 
searching over a series of fine steps in $g,g-r,$ and $\rm DM$ values for each isochrone.  Input isochrones are 
supplied by the Dartmouth library \citep{Dotter08}, and linearly interpolated at a step size of 0.01 mag in the 2D 
color--magnitude space.  The distance modulus is sampled over a range of $15.5 < \rm m-M < 18.5$ 
in steps of 0.025 mag. To achieve the best sensitivity in the fitting process we used the isochrone segment between 
0.5\,mag brighter than the MSTO and 2\,mag below. 

\begin{figure*}\centering  
\includegraphics[width=6.2cm]{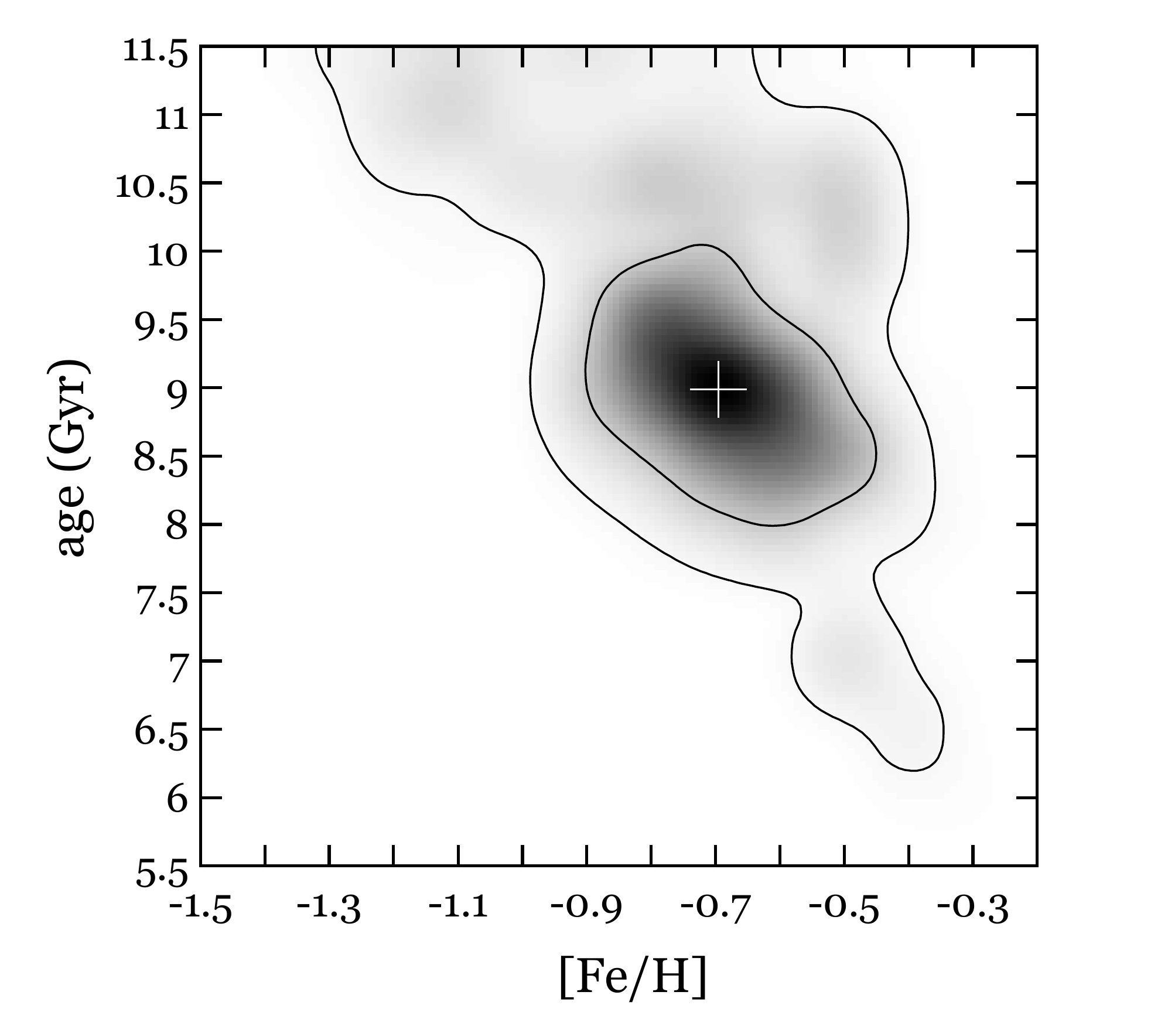}\hspace*{-0.5cm}
\includegraphics[width=6.2cm]{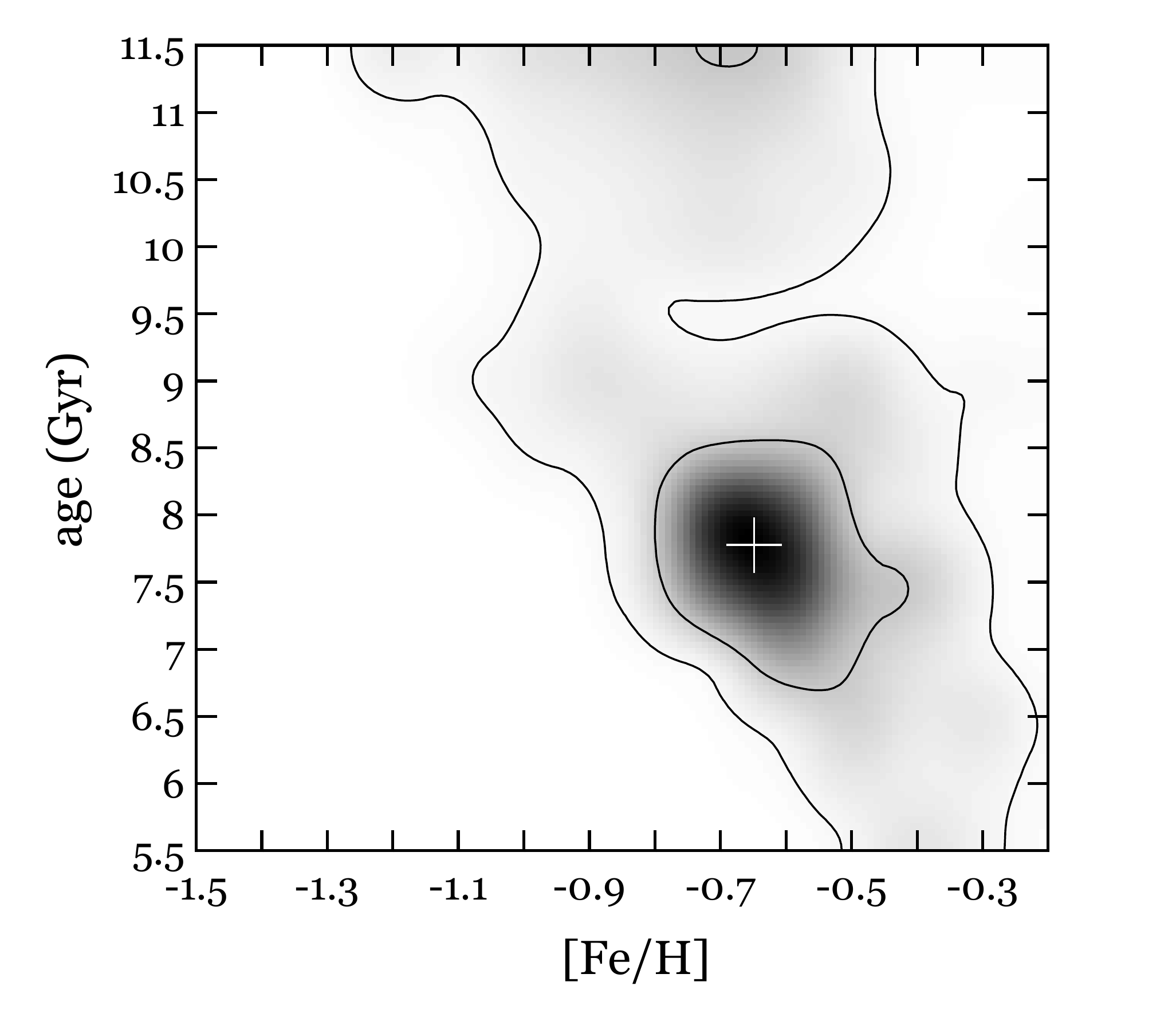}\hspace*{-0.5cm}
\includegraphics[width=6.2cm]{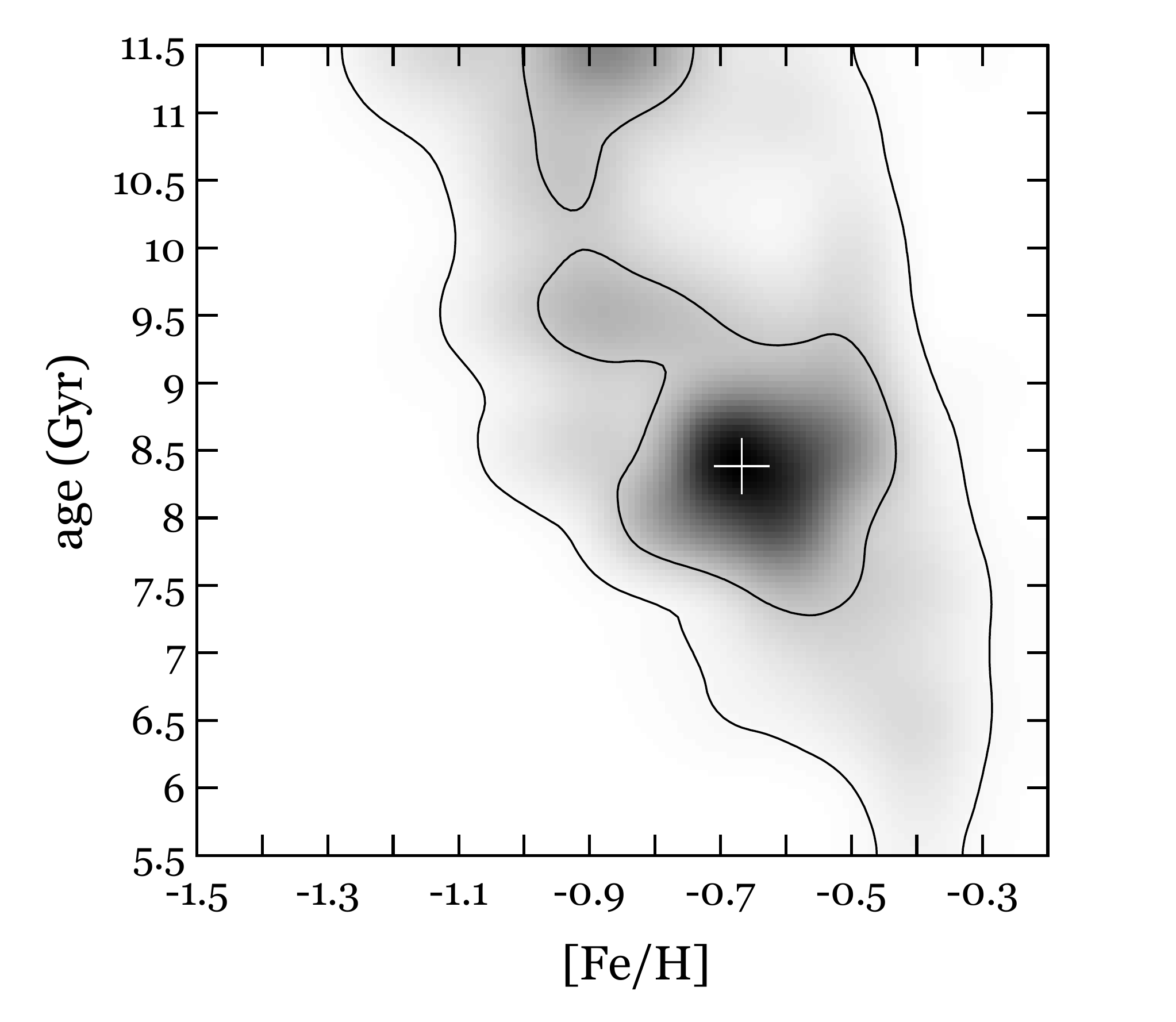}
\caption{Smoothed maximum likelihood solutions in age-abundance space for the main sequences in the $1145+13$ (left), 
$1220-01$ (middle) and $1247-00$ (right) fields with contours at the 68\% and 95\% levels.
Note that for all panels, any fit at relatively low abundances, e.g.\ $-1.5$, and at
reasonable ages ($<11.5$\,Gyr) is well outside the 2$\sigma$ boundary.
\label{ml_solutions}}\label{ML_solutions}
\end{figure*}

We calculate the maximum likelihood values $\mathcal{L}_i$ over a grid of isochrones, covering an age range from 5.5 
to 13.5\,Gyr and metallicity range $-2.5\leq$ [Fe/H] $\leq-0.2$\,dex.  Grid steps are 0.5\,Gyr in age, and 0.1\,dex in [Fe/H].  
With a grid of $\mathcal{L}_i$ values, we can locate the most likely value and compute confidence intervals by 
interpolating between grid points.  In addition to this interpolation, we smooth the likelihood values over $\sim2$ grid 
points in order to provide a more conservative estimate of parameter uncertainties.  In Figure \ref{ML_solutions}, we 
present the relative density of likelihood values for the sample described above.  We find the isochrone with the 
highest probability has an age of $9.1$\,Gyr  and [Fe/H] $=-0.70$, with 68\% and 95\% confidence contours presented in 
the figure. The marginalized uncertainties (Table\,\ref{MS_paras}) about this most probable location correspond to an age of 
$9.1^{+1.0}_{-1.1}$\,Gyr, a metallicity of [Fe/H]$=-0.70^{+0.15}_{-0.20}$\,dex, and a distance modulus of 
$\rm DM=17.45\pm0.20$\,mag ($d=30.9\pm3.0$\,kpc). They account for the varying photometric errors in the critical 
part $20<g<23$ of the CMD, the differences in Dartmouth isochrone shapes between 5-12\,Gyr, and isochrone model 
variations. Other sources of uncertainties such as the systematic errors inherent in the process of isochrone fitting,
small errors in the photometric zero points and, at a more fundamental level, the underlying stellar physics and stellar 
populations used to calibrate the theoretical isochrones mean that the error bars are in general underestimates. 
Nevertheless, the conclusions we can draw from the results are unaffected.

The best-fitting isochrone is overplotted as a blue line in the left panel of Fig.~\ref{cmdfits}. 
The associated heliocentric distance of the main sequence of $30.9\pm3.0$\,kpc is in excellent 
agreement with the estimated Sgr Stream distance of 29\,kpc at that location \citep{Belokurov06b}. 
To illustrate that our results are independent of the chosen set of theoretical models we also 
present the Padova isochrone \citep{Girardi04} with the same age, [Fe/H], and distance
(red curve).
The slightly brighter turnoff and a slightly bluer lower giant branch corresponds to a metallicity 
difference of [Fe/H]$_D$--[Fe/H]$_P=0.07$\,dex and an age difference of $0.7$\,Gyr, well 
within the listed uncertainties derived from the Dartmouth isochrones.
For a detailed comparison of the Dartmouth and Padova isochrones we refer to \cite{Dotter07, Dotter08},
but we note that the differences for low mass stars is because of differences in the treatment of the equation 
of state and surface boundary conditions. These differences are additional reasons why isochrone fitting was restricted to 
the brighter parts of the main sequence.

There is not much information currently available in the literature about the age, metallicity and color 
of the leading tidal tail MSTO in this part of the Sgr Stream. \cite{Carlin12} presented the most recent results for fields 
in the trailing tidal tail.  Their Fig.~21 shows the spectroscopic MDFs of four fields along the regime $75^\circ<\Lambda_\sun<130^\circ$ 
with no significant metallicity gradient observed within the uncertainties. From the numbers in their Table\,6 we calculate a mean 
[Fe/H] of $-1.15$ and a mean $\sigma_{\rm [Fe/H]}$ of 0.64. That $1\sigma$ range of $-1.77<$[Fe/H]$<-0.49$ is statistically in 
agreement with our quoted $1\sigma$ range of $-0.90<$[Fe/H]$<-0.55$. 

\begin{table}\centering
\caption{Properties of the three Overdensities}
\begin{tabular}{lcccc}
 & m-M & $\langle D\rangle_{\odot}$ & Age & [Fe/H]  \\
Field & (mag) & (kpc) & (Gyr) &  (dex) \\[2pt]
\hline\\[-8pt]
1145$+$13    & $17.45\pm0.20$   & $30.9\pm3.0$ & $9.1^{+1.0}_{-1.1}$& $-0.70^{+0.15}_{-0.20}$\\[2pt]
1220$-$01 &  $16.93\pm0.21$  & $24.3\pm2.5$ & $7.9^{+0.7}_{-1.2}$ & $-0.66^{+0.20}_{-0.14}$\\[2pt]            
1247$-$00 &  $16.73\pm0.19$  & $22.2\pm2.1$ & $8.5^{+1.5}_{-1.2}$ & $-0.68^{+0.24}_{-0.20}$\\[2pt]
\hline\\ \label{MS_paras}
\end{tabular}
\vspace{-0.5cm}
\tablecomments{Parameters inferred from fitting Dartmouth isochrones to the main sequence.}
\end{table}

\subsection{The VSS/VOD Fields}
The detection algorithm used by \cite{Walsh09} to search for new ultra-faint Milky Way dwarf 
satellites flagged two positions, $1220-01$ and $1247-00$ (see Table\,\ref{field_coords}) both 
located in the vicinity of two fields studied by \cite{MartinezDelgado07} and 
in the general direction of the Virgo Stellar Stream/Virgo Overdensity as defined by excesses of 
RRL stars \citep{Vivas01, Vivas02, Vivas03, Zinn04, Ivezic05} and of F-type main sequence stars
\citep{Newberg02}, spanning an R.A./Decl.~range of $175^\circ<\alpha<200^\circ$ and  $-2.3^\circ<\delta<0.0^\circ$.
The reddening corrected CMDs of $1220-01$ and $1247-00$  are shown in 
Fig.~\ref{rawcmds}. 
Main sequences are clearly visible in the field-subtracted CMD from g$\sim$20 down 
to 24.0\,mag, almost as conspicuous as in our Sgr Stream field. Following the same procedure 
as described before we established the best-fitting isochrones. They are shown in Fig.~\ref{cmdfits}
with the corresponding maximum likelihood solutions in Fig.~\ref{ML_solutions}. We derive
similar ages of $7.9^{+0.7}_{-1.2}$\,Gyr and $8.5^{+1.5}_{-1.2}$\,Gyr
and identical metallicities of [Fe/H]=$-0.66^{+0.20}_{-0.14}$
and  $-0.68^{+0.24}_{-0.20}$, respectively. 
The associated heliocentric distances are $24.3\pm2.5$\,kpc  and $22.2\pm2.1$ (see Table\,\ref{MS_paras}).

We estimate the strength of the main sequence population in the two VOD fields from star counts 
within $|g-r|<0.2$\,mag of the best-fitting isochrone and 3\,mag down the MS from 
the turn-off. These numbers are listed in Table\,\ref{paras} together with the corresponding percentages 
normalised to the 249 stars found in the $1145+13$ Sgr Stream field. The quoted uncertainties are the 
$\sqrt{n}$ standard error. Although the VOD fields are $\approx 13^\circ$ away from the main ridge 
of the Sgr Tidal Stream we observe only a moderate drop in MS star counts to about 50\,percent. 

\begin{figure}[ht]  %
\includegraphics[width=4.7cm]{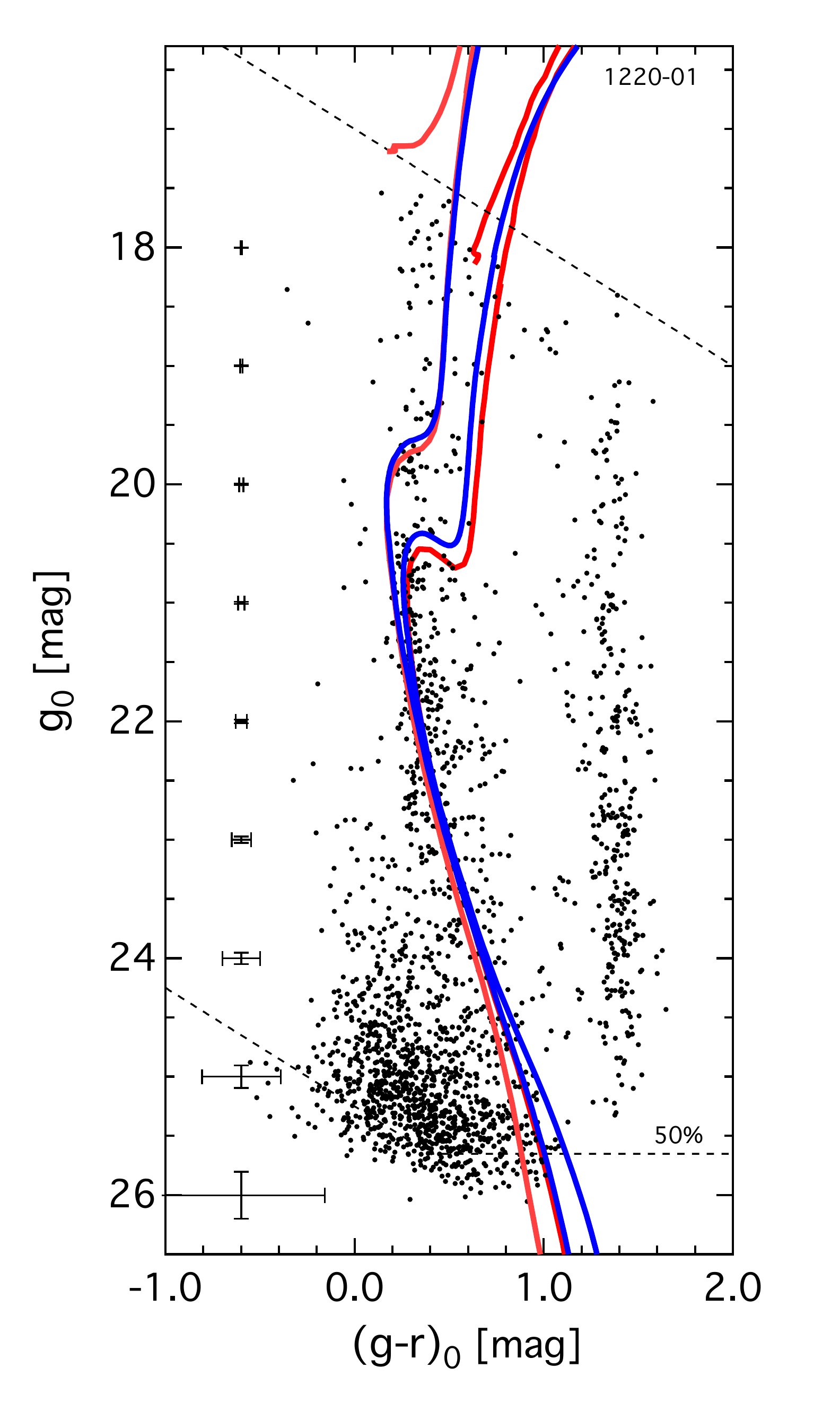}\hspace*{-0.5cm}
\includegraphics[width=4.7cm]{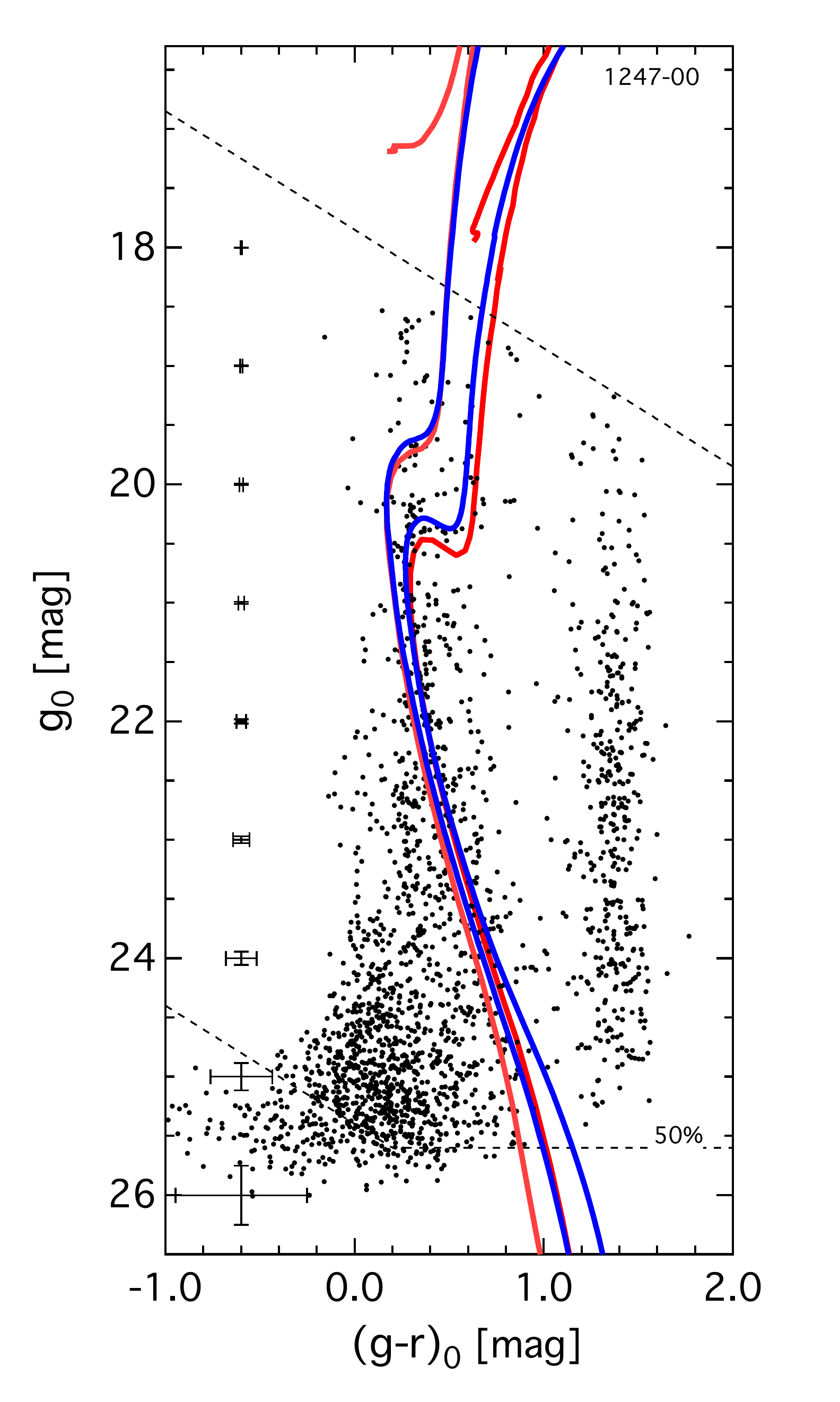}\hspace*{-0.5cm}
\caption{The same CMDs of the 1220$-$01 and 1247$-$00 fields as in Fig.\ref{cmdfits}. 
The Dartmouth and Padova isochrones of an old (age=11.2\,Gyr), metal poor ([Fe/H]=$-1.9$)
stellar population at 19\,kpc as prescribed by the RRL stars in the VSS/VOD region 
are superimposed on the CMDs shown in Fig.\ref{cmdfits} as blue and red lines, respectively. 
In both fields, the MSTO region matches a small group of stars at $g\approx19.7$.
However, VSS/VOD-associated main sequence stars are mostly absent for about $0.7-1$\,mag 
below the turn-off. \label{iso_comparison}}
\end{figure}

\cite{Chou10} pointed out that the lack of metal-rich RRL stars could mean that 
the entire VSS/VOD stellar population is metal-poor. With this point as a base, we 
investigate the possible presence of an old, metal-poor stellar population in the 1220$-$01
and 1247$-$00 CMDs. For that purpose, we take the estimates of $-1.86$ to $-2.0$ 
for the metallicity of VSS/VOD RRL stars \citep{Duffau06, Prior09b, An09}, 
adopt an age  of 11.2\,Gyr, and distance of 19\,kpc \citep{Duffau06, Newberg07, Bell08, Prior09b} 
and overplot in the CMDs the corresponding Dartmouth and Padova isochrones (Fig.~\ref{iso_comparison}). 
In both CMDs, the MSTO region of the old, metal-poor isochrones seems to match a small group of stars 
at $g\approx19.7$. However, the bright main sequence stars that would go with this feature
are mostly absent for about $0.7-1.0$\,mag 
below the turn-off until the two isochrones merge with the best-fitting isochrones. The lack of such
bright main sequence stars in the interval $20<g<20.8$ that could be associated to the VSS/VOD 
suggests that the nearer population, i.e.\ at ~19\,kpc, is small at best when compared to the slightly 
more distant main sequence stars at $23$\,kpc.

To quantitatively assess the possibility that some of the stars above the MSTO ($g<20.2$ in Fig.~\ref{cmdfits}) 
are from a smaller VOD stellar population located a few kpc closer along the line-of-sight we count stars around the 
MSTO with colors $0.2<g-r<0.5$ in the decontaminated CMDs. The number of stars within one 
magnitude above and below the MSTO are listed in Table\,\ref{starstats}. The 1220$-$01 and 
1247$-$00 fields have marginally more stars ($50\pm15$\%) above the turn-off when compared to 
the 1145+13 field ($34\pm9$\%).  If approximately 1/3 of the stars above the MSTO are 
associated to the Milky Way (i.e.~leftover stars from the decontamination process), then the 
metal-poor, old VOD population is about 20-30\% compared to the VOD main sequence stars in that part of the sky.

We now discuss possible interpretations of the results from the CMD analysis.

\section{Are the VOD main sequence stars part of the Sgr tidal Stream?}
Intriguingly, the statistically robust results, based on  about 260 main 
sequence stars ($0<g-r<1$, $g<24$) in the two VOD fields, 
are in excellent agreement with the age and metallicity derived for the MS stars 
in the 1145+13 field selected to be in the direction of the highest star density of the 
Sgr Branch\,A leading arm. This immediately raises the question, are the detected 
VOD main sequence stars part of the Sgr Tidal Stream? We find that the VOD results are in good 
agreement with the age for the main population in the Sagittarius dwarf \citep[$8.0\pm1.5$\,Gyr;][]{Bellazzini06} 
and  the peak in the metallicity distribution function at [Fe/H]=$-0.7$\,dex for the $2-3$\,Gyr 
old M-giants in the Sgr north leading arm in the $\Lambda_\sun=260^\circ$ region \citep{Chou07}.
We recognise that the M-giants might be biased to higher metallicities compared to a complete RGB sample
but the Bellazzini et al. study is principally based on K-giants and does not indicate any major difference with the M-giant results. We also 
note that the metallicity of the VOD MS stars is slightly lower when compared to the 
stars in the Sgr core where the metallicity distribution function has a wide spread from $-1.0$ to super-solar $+0.2$ and 
peaks at [Fe/H]=$-0.4$\,dex \citep{Bellazzini08}. However, this difference can be explained with the reported abundance 
gradient along the tidal arms \citep{Keller10}.

\begin{table}
\begin{center}
\caption{Star Counts for the Three Main Sequences}
\begin{tabular}{lcr}
Field& Counts & Strength  \\\hline
1145$+13$ & 249& 100\% \\
1220$-$01 & 134 & 54$\pm7$\% \\
1247$-$00 & 132 & 53$\pm7$\% \\\hline
\label{paras}
\end{tabular}
\end{center}
\vspace{-0.5cm}
\tablecomments{Star counts within $|g-r|<0.2$\,mag of the best-fitting isochrone and 3\,mag 
down the MS from the turn-off. The third column gives the population strength relative to the main 
sequence in Sgr Stream Branch A.}\hfill
\end{table}

\begin{table}
\begin{center}
\caption{Number statistics of stars around the MSTO}
\begin{tabular}{lccl}
Field& Above& Below& Percent  \\\hline
1145+13 & 33 & 97& $34\pm9$\% \\
1220$-$01 & 38 & 82 &$46\pm13$\% \\
1247$-$00  & 33 & 60 & $55\pm17$\%\\\hline
\label{starstats}
\end{tabular}
\end{center}
\vspace{-0.5cm}
\tablecomments{Stars are counted in the $0.2<g-r<0.5$ color interval and within one magnitude above and below the MSTO. The fourth 
column gives the number ratios with $\sqrt{n}$ uncertainties.}
\end{table}

Consequently, based solely on the similarity between the MS population parameters derived for our VSS/VOD fields and those observed for the Sgr dwarf and the Sgr Tidal Stream stars, it would be possible to draw the conclusion that the stellar population of the complex overdense region around 
$\alpha$=12.5\,h, $\delta$=0\,deg in the Virgo constellation is dominated by Sgr Tidal Stream main sequence stars.  However, such a conclusion ignores the additional information from models of the Sgr/Galaxy interaction, which provide further constraints.  These will be discussed in the next section.  We note, however, that, as discussed in Section 3.2, regardless of the origin of the dominant population, our data provide little evidence for the presence of a significant separate old and metal-poor population of stars at a distance of $\sim19$\,kpc.

\begin{figure}[t]  
\centering
\includegraphics[width=8.1cm]{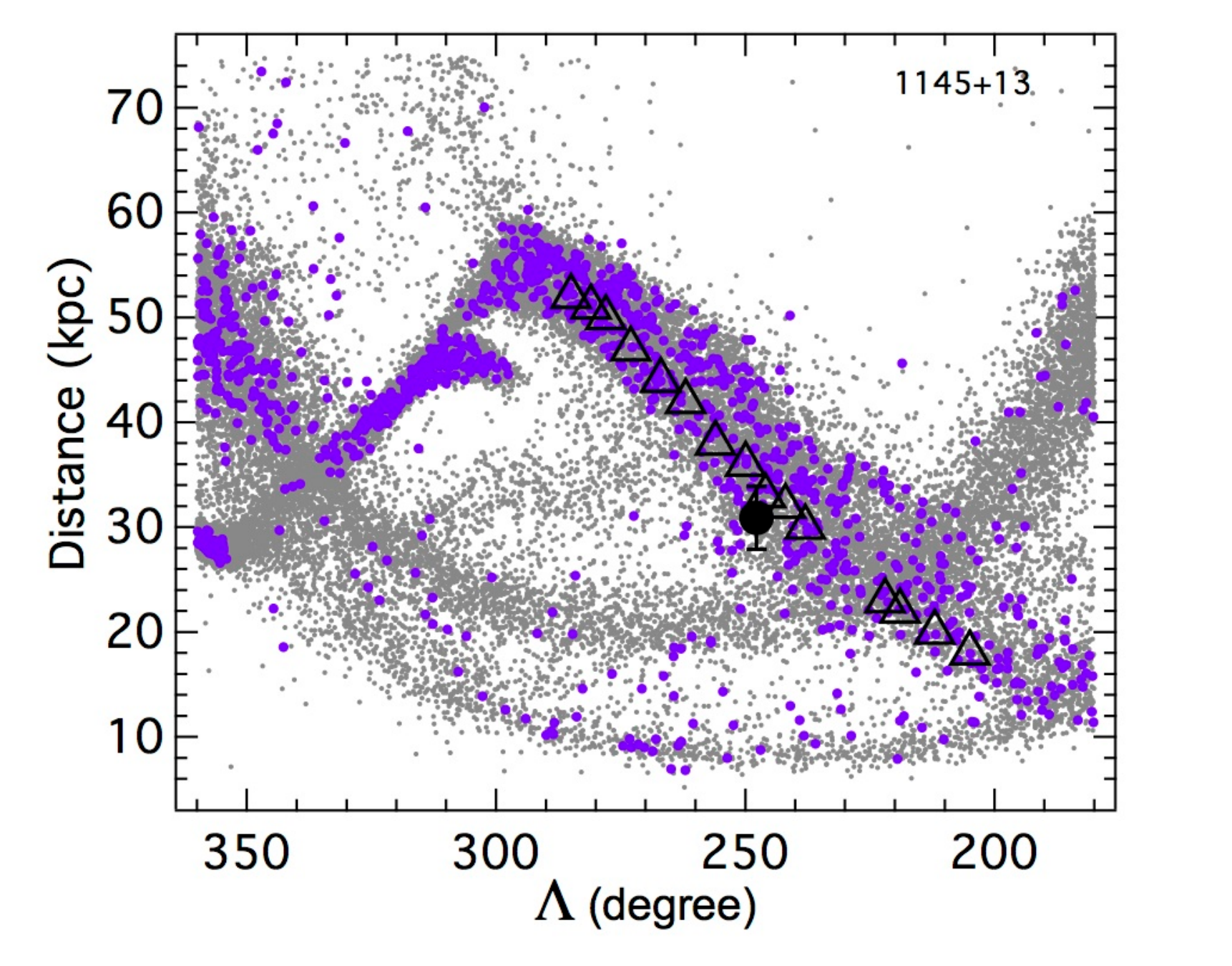}\vspace{-.2cm}   
\includegraphics[width=8.1cm]{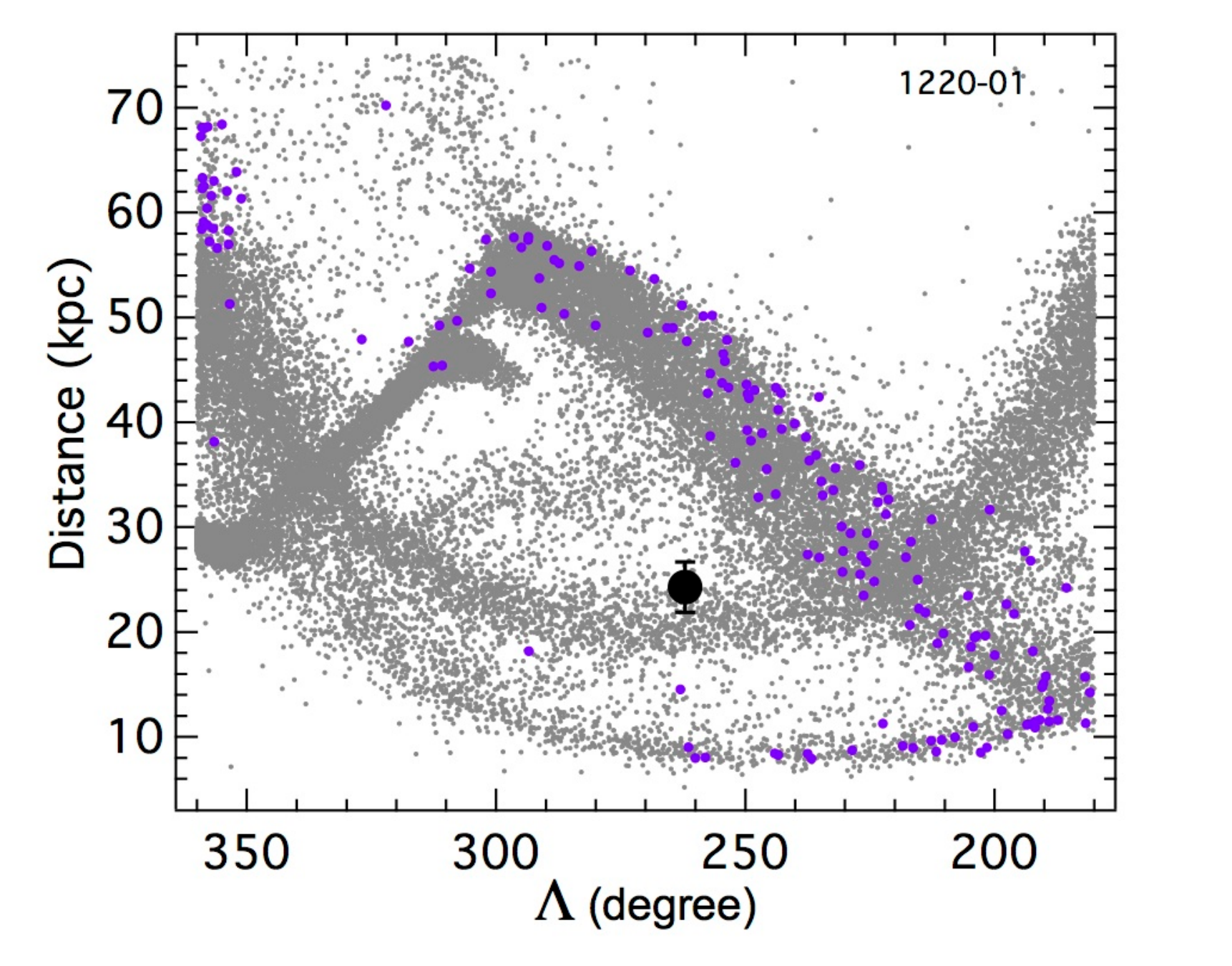}\vspace{-.2cm}      
\includegraphics[width=8.1cm]{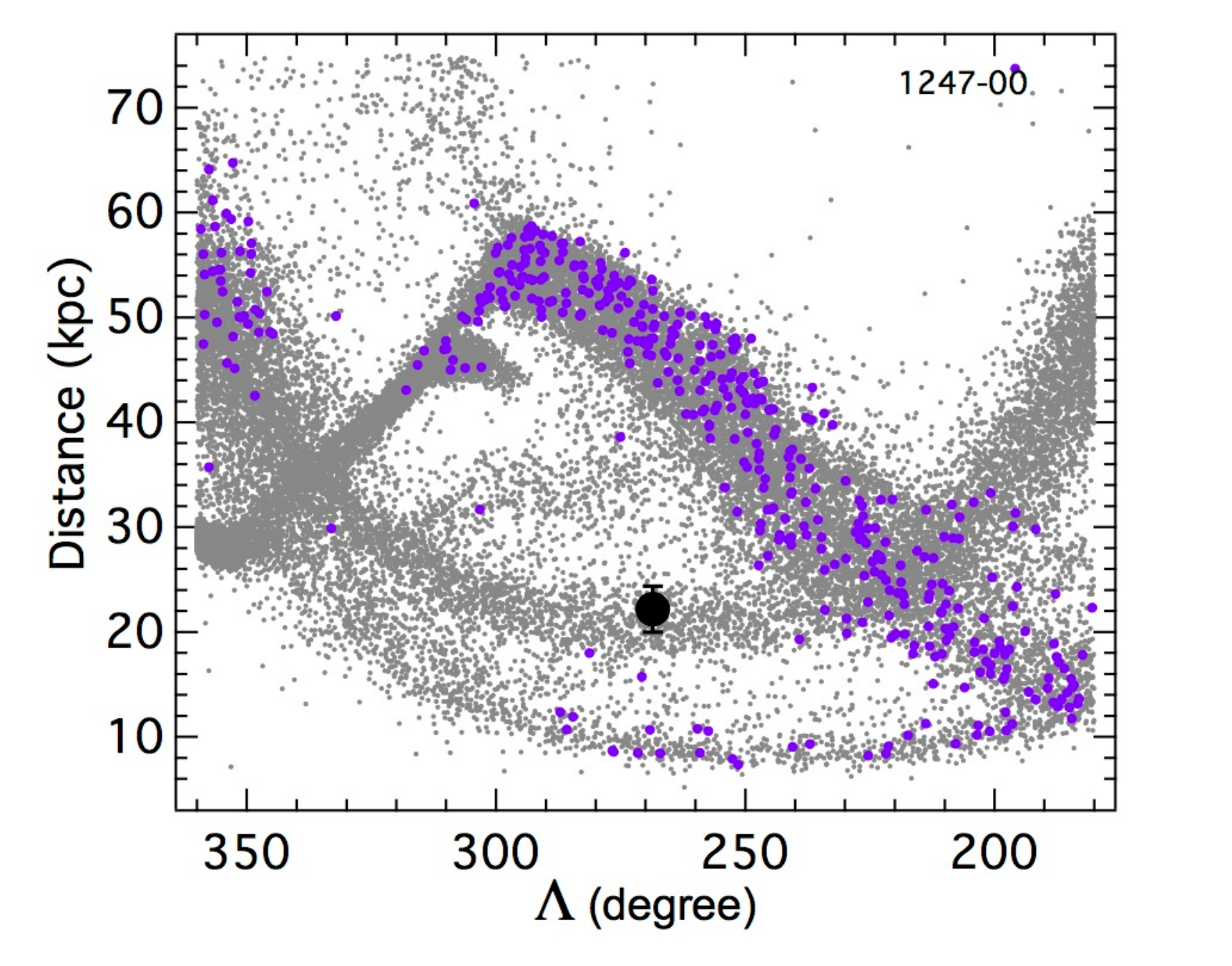} \vspace{-.2cm}     
\caption{ Distribution of the LM10 model particles in the $180^\circ<\Lambda_\odot<360^\circ$ range with 
the points in the orbital latitude interval $4.95^\circ<B_\odot<5.45^\circ$ (top), $15.05^\circ<B_\odot<15.55^\circ$ (centre)
and $11.15^\circ<B_\odot<11.65^\circ$ (bottom) highlighted in purple. The triaxial model agrees well with the 
spatial location of the MS stars in the Sgr Tidal Stream field $1145+13$ (filled circle) and the SDSS A-stream data (triangles)
from \cite{Belokurov06b}. For the two VOD fields the model predicts a small concentration of
Sgr Tidal Stream stars at 9\,kpc and in the distance intervals $46<D<54$\,kpc and $42<D<53$\,kpc, respectively, but no particles at 23\,kpc where 
the most prominent population of main sequence stars is detected. From particle number counts we estimate 
that the Sgr Tidal Stream stars in the background should be more abundant  in the $1247-00$ 
field than in the $1220-01$ field by approximately a factor of 4.3.
\label{Law10}}
\end{figure}

Finally, we like to point out that the RRLs probably associated with Sgr in this region do not pose a problem. 
Abundances for 14 RRLs 
where the association with Sgr is unambiguous \citep[e.g.][]{Vivas05} are low 
($\langle$[Fe/H]$\rangle= -1.76$ with $\sigma=0.22$). Moreover, the six \cite{Starkenburg09} 
Group 1 stars, also likely Sgr objects, lie in this same part of the sky and have a similar mean abundance.
This abundance is consistent with the RRLs in the VSS/VOD region (whether they are VSS objects or not).
For instance, the two large negative V$_{GSR}$ RRLs in \cite{Prior09a}, which are likely Sgr objects, have 
abundances $\langle$[Fe/H]$\rangle=-1.57$ similar to the Sgr RRLs of \cite{Prior09b} and those in the 
Vivas et al.~paper.  In other words, we are not trying to deny that there is a metal-poor Sgr population, 
only that it is not dominant and that the mean abundance of Sgr RRL stars is a poor indicator of the mean 
abundance of the Sgr Stream, as the RRLs are likely older than the bulk of the Sgr stellar population.

\section{Are the VOD main sequence stars from a different halo structure?}
From the analysis of the deep CMDs we also gained information 
about the 3-D location of significant populations of main sequence stars in three well-defined directions. 
These geometrical anchor points allow a comparison with the currently best Sgr 
Tidal Stream model by \citet[hereafter LM10]{Law10}.
The N-body simulation 
from Law \& Majewski is based on a complete all-sky view of the Sgr Stream. It is empirically 
calibrated with SDSS A-branch and 2MASS M\,giant stars and predicts the 
heliocentric distances and radial velocities of 100,000 particles stripped from the Sagittarius dwarf up to four 
orbits (approximately 3\,Gyr) ago. 

In Fig.~\ref{Law10} we plot heliocentric distance data as a function of orbital longitude for the simulated 
Sgr satellite debris from the best-fit triaxial model ($a:b:c= 1:0.99:0.72$) within the assumptions of the 
simulation\footnote{http://www.astro.virginia.edu/$\sim$srm4n/Sgr/data.html}. 
The locations of the MS stars in our three fields are shown as filled circles. To illustrate 
the model predictions we highlight those Sgr Stream model particles 
that fall within half a degree of each field's orbital latitude. The chosen width equals the observed field-of-view. 
The distance distribution of the model particles is bimodal in all three cases 
with a minor concentration at 9\,kpc and a slightly larger population between $46<D<54$\,kpc and $42<D<53$\,kpc, 
respectively. The model prediction agrees well with the distance we obtained from isochrone fitting for the MS stars in the 1145$+$13 field 
(top panel of Fig.~\ref{Law10}). The good match was in fact anticipated as the LM10 Sgr Tidal Stream model has been 
calibrated with the SDSS A-branch data from \cite{Belokurov06b} (triangles in Fig.~\ref{Law10}). However, no agreement 
is found between the model and the VOD main sequence stars (filled circle in centre and bottom panels of Fig.~\ref{Law10}): no Sgr Tidal Stream 
stars are expected at the distance of 23\,kpc where the maximum-likelihood fit located the main sequence in the 
1220$-$01 and 1247$-$00 fields. 
If we adopt the LM10 triaxial model as the true representation of the Sgr Tidal Stream in this part of the sky then 
the necessary conclusion would be that the observed MS stars at 23\,kpc originate from a different
stellar structure in the Milky Way halo.

\begin{figure*}  
\centering
\includegraphics[width=5.7cm]{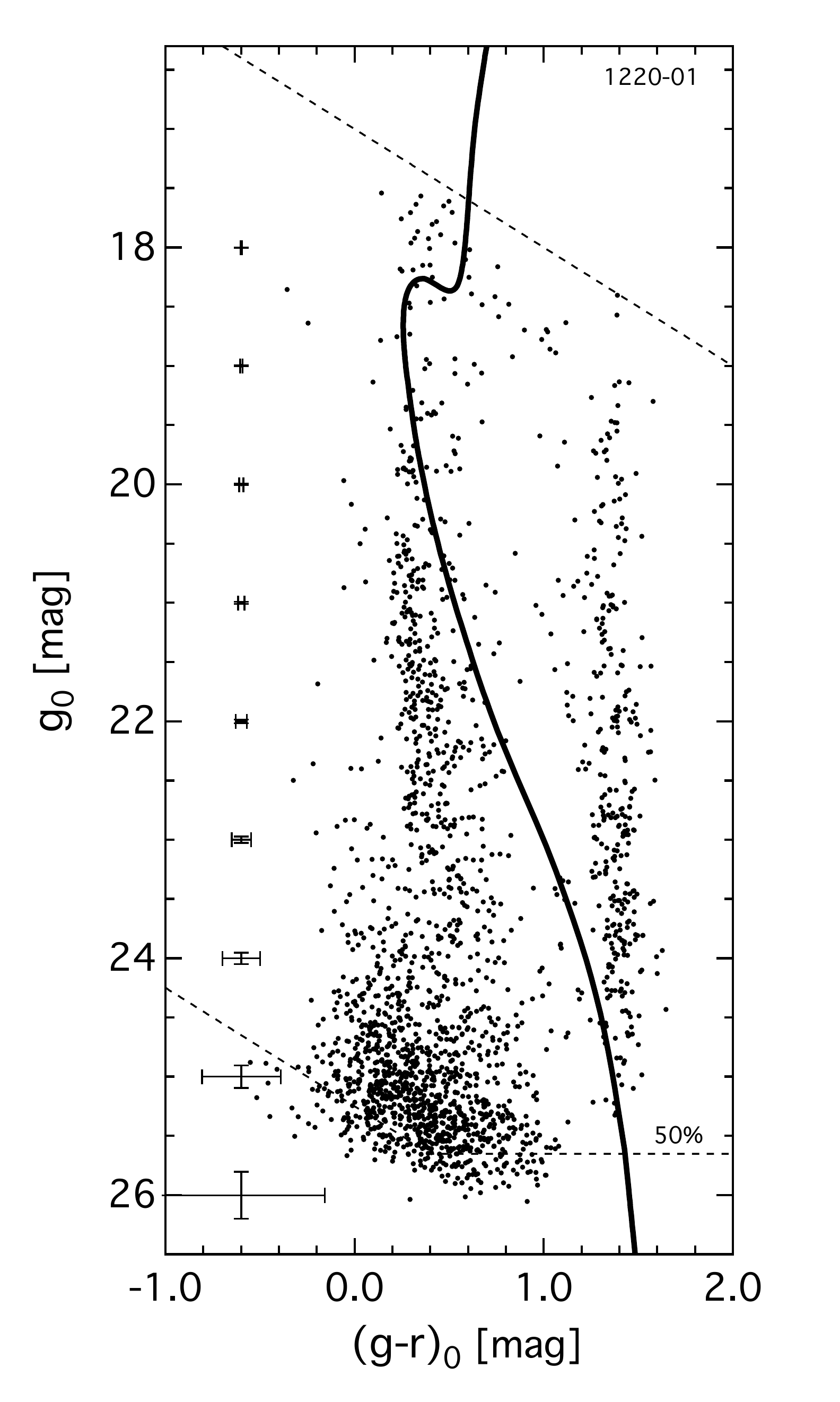}\hspace*{-0.5cm}   
\includegraphics[width=5.7cm]{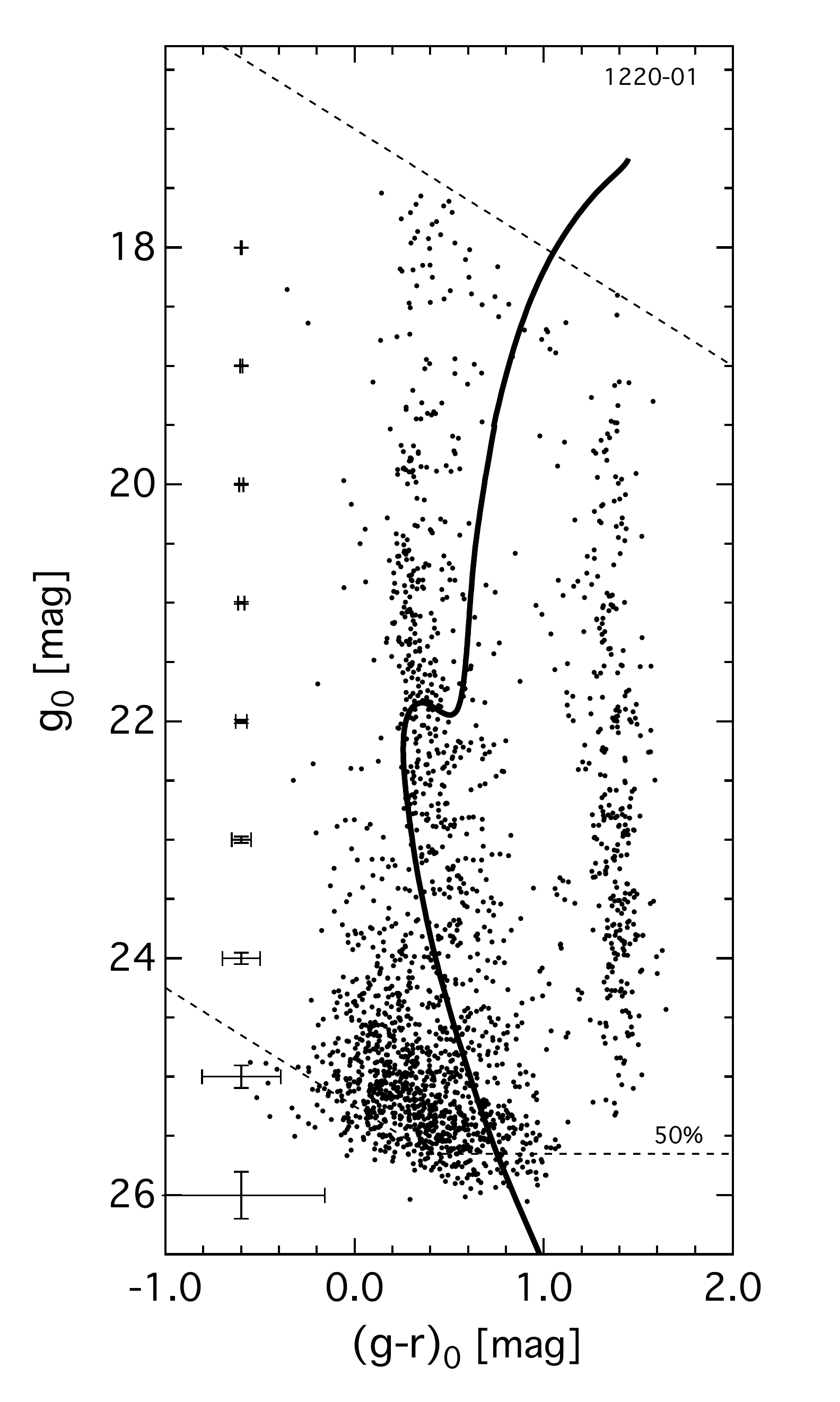}\\
\includegraphics[width=5.7cm]{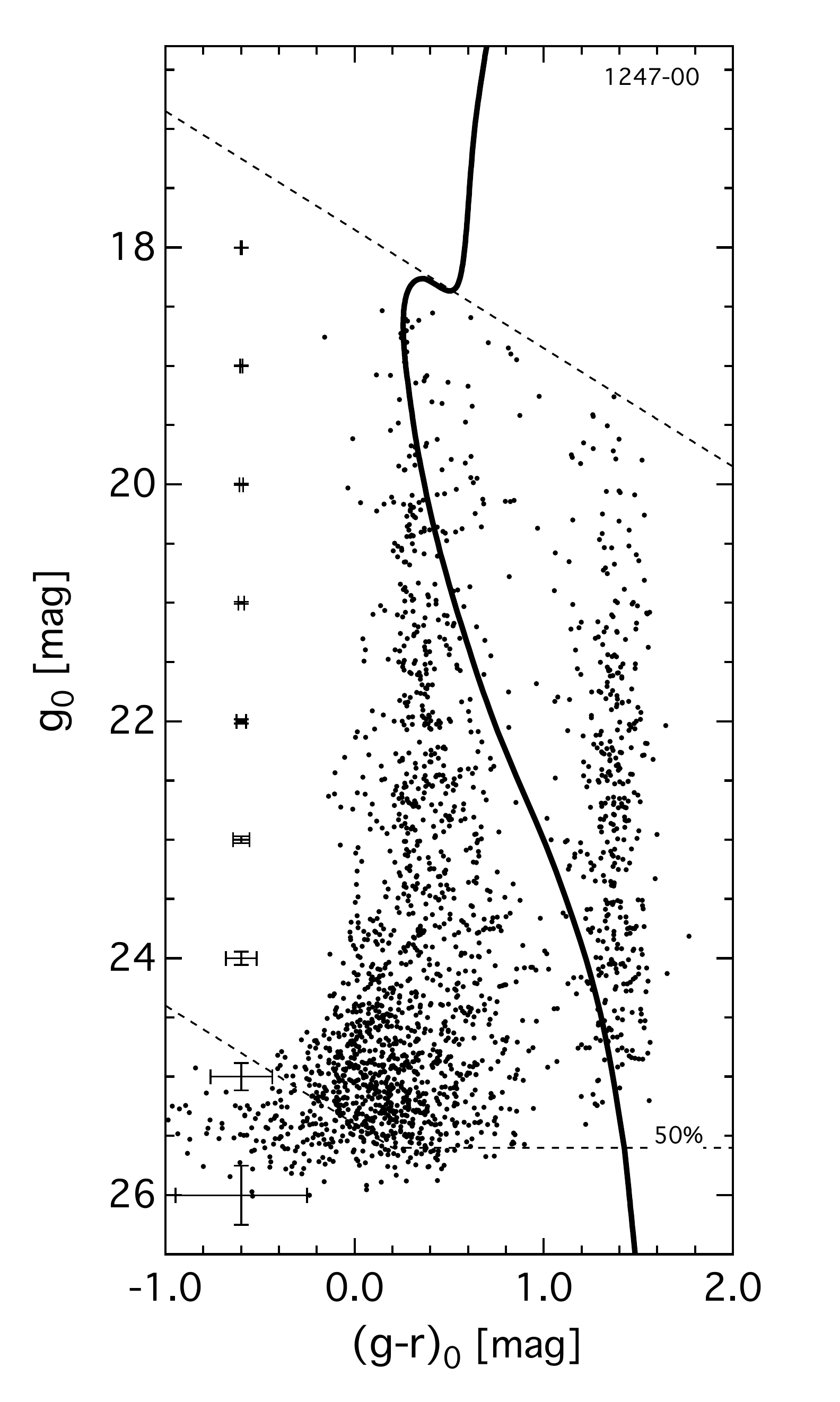} \hspace*{-0.5cm} 
\includegraphics[width=5.7cm]{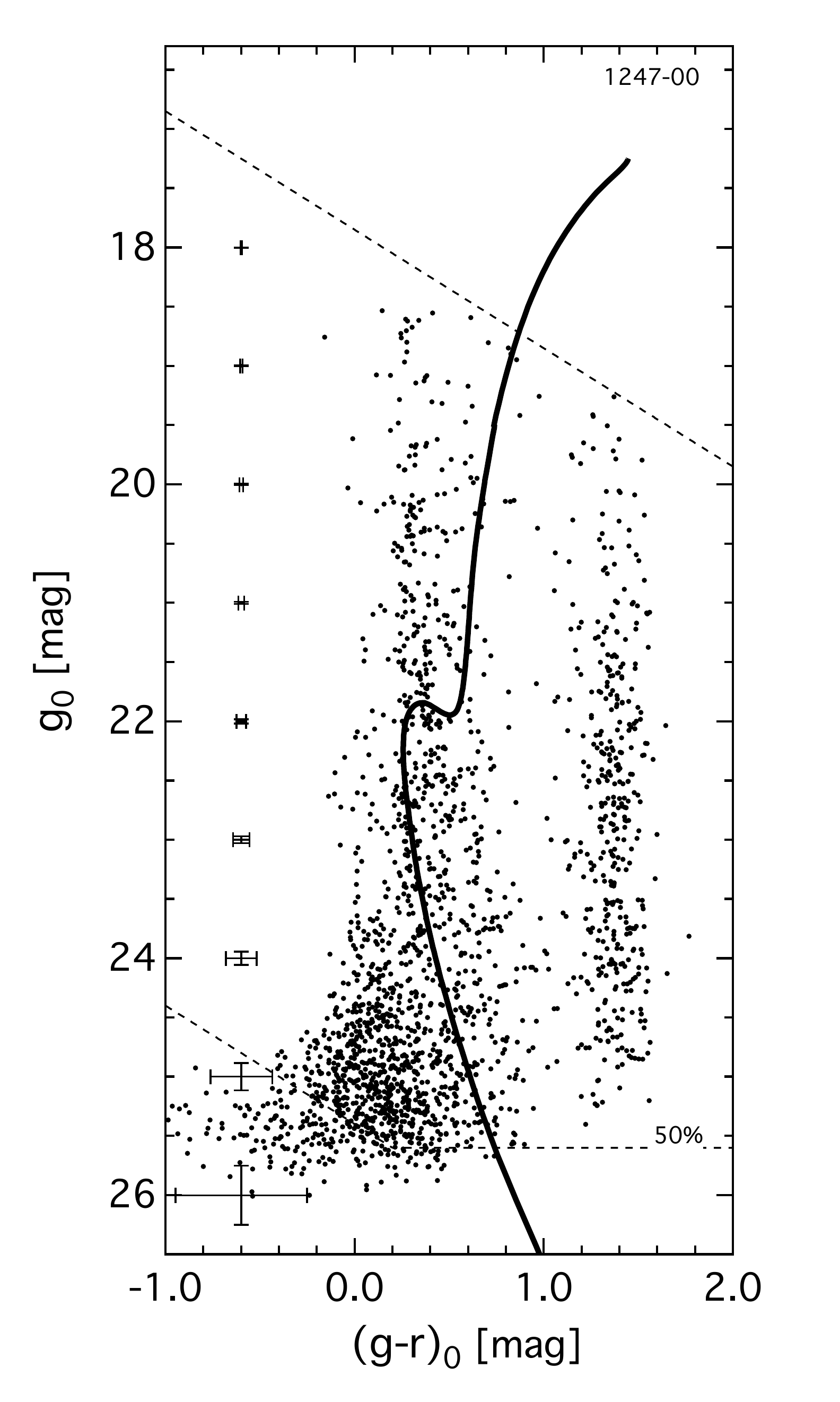}\\
\caption{Dartmouth isochrones superimposed over the $1220-01$ (top row) and
$1247-00$ CMD (bottom row) representing stellar populations at the distance of 
9\,kpc and 47\,kpc, respectively. The isochrone at 9\,kpc  (left column) has a metallicity of 
[Fe/H]=$-0.70$ and an age of 8\,Gyr. In both CMDs we notice groups of stars that follow 
the isochrone resembling the small Sgr Tidal Stream population at 9\,kpc predicted by the LM10 model. This notion
finds further support from the filamentous structure seen in the combined CMD (top panel, Fig.~\ref{9kpc_feature}). The panels 
on the right show the same isochrone at a distance of 47\,kpc. Stars are scattered differently around the isochrone in the two fields,
indicative for a larger population of Sgr Sream stars in the $1247-00$ field, as is also predicted by the model.\label{distrange}}
\end{figure*}

\begin{figure}  
\centering
\includegraphics[width=6.7cm]{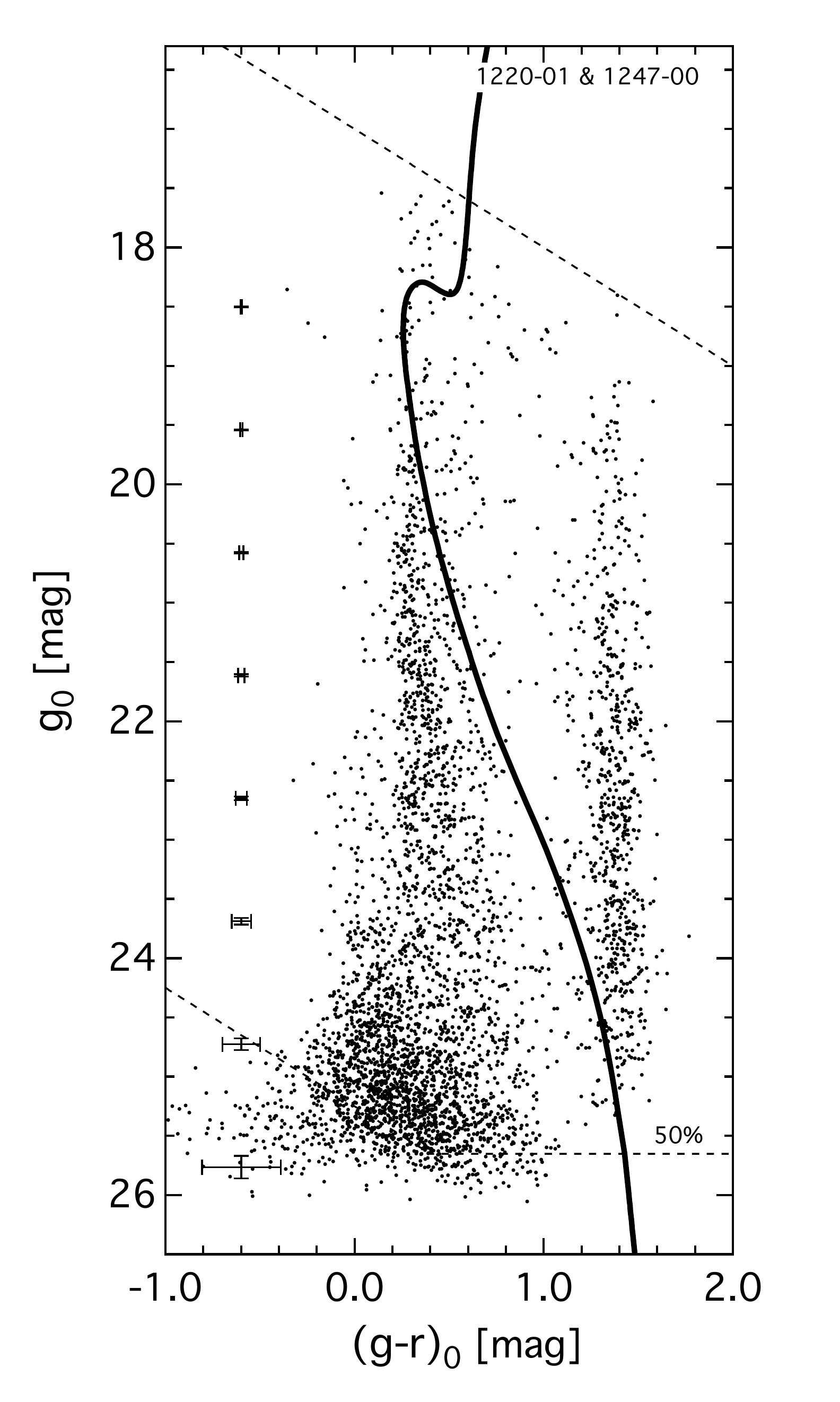}\vspace{-0.5cm}\\
\includegraphics[width=6.7cm]{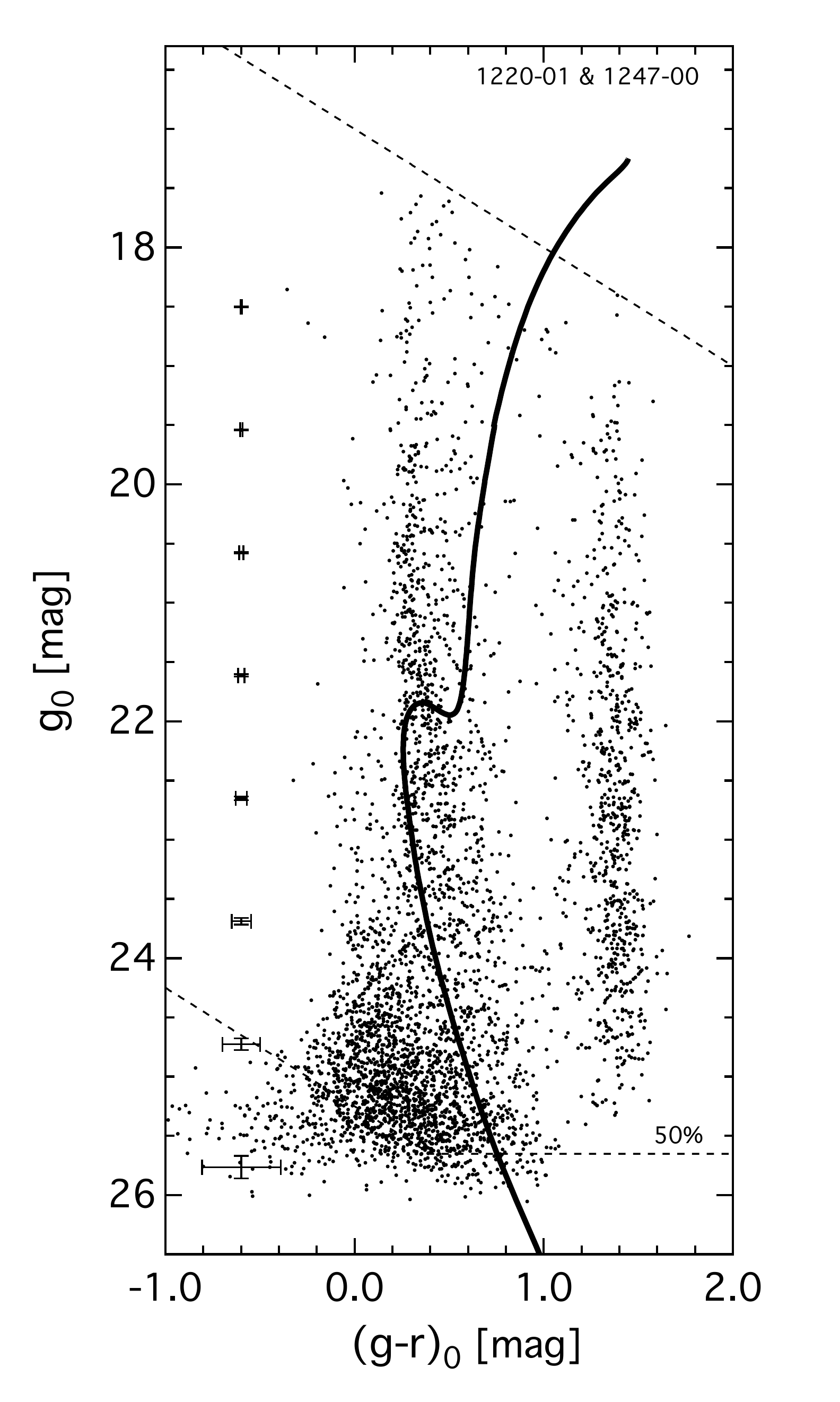}
\caption{Same as left and center panels in Fig.\ref{distrange} but for the combined CMDs. (Top)
we notice a distinct narrow feature below the isochrone at a distance of about 10\,kpc following closely 
the isochrone ($21.2<g<22.5$, $0.4<g-r<0.8$). In terms of distance, small distance spread and 
population strength this feature could well be the 9\,kpc Sgr Tidal Stream predicted by the LM10 model. 
(Bottom) It is plausible that the stars scattered around the 47\,kpc isochrone are from the
 Sgr Tidal Stream that extends from $42<D<53$\,kpc along the line-of-sight.
\label{9kpc_feature}}
\end{figure}

Do we find traces of Sgr Stream stars in our CMDs? To guide 
the eye we superimpose a Dartmouth isochrone with a metallicity of [Fe/H]=$-0.70$ and an age of 8\,Gyr
at distances of 9 and 47\,kpc over the 1220$-$01 and 1247$-$00 CMDs in the panels of Fig.~\ref{distrange}.
Looking at the course of the isochrone in both lefthand panels we notice a diagonal ensemble of stars following 
the isochrone and running in parallel to the main sequence at 23\,kpc from $g_0=18.5$\,mag down to about $g_0=22.5$\,mag. 
It is conceivable that these 
stars are indeed members of the small population of Sgr Stream stars at 9\,kpc as predicted by the model. To enhance the signal 
we combine the CMDs of the two fields in Fig.~\ref{9kpc_feature}. Now we can see a distinct narrow feature 
($20.5<g_0<22.5$, $0.4<(g-r)_0<0.8$) parallel and close to the isochrone at a distance of approximately 10\,kpc. In terms of 
distance, small distance spread and population size this filamentary structure is suggestive of the 9\,kpc Sgr Tidal Stream  
predicted by the LM10 model. We note that this view offers an interesting and alternative interpretation of the origin of stars brighter than 
the 23\,kpc MSTO as it has been discussed in Section 4.1. They are main sequence stars of the Sgr Stream at 9\,kpc 
rather than the MSTO stars of an old, metal-poor population at 19\,kpc.

Moving on to the 47\,kpc feature (righthand panels in Fig.\,\ref{distrange}) reveals a difference between the 1220$-$01 and
1247$-$00 CMDs. In the latter field we observe more stars scattered around the isochrone 
while the fewer stars in the 1220$-$01 CMD close to the isochrone lie systematically above the line. 
This finding is again consistent with the picture we get from the LM10 
model (see Fig.~\ref{Law10}) that predicts 
a population of Sgr Stream stars in the 1247$-$00 direction larger by a factor of 4.3. Hence it is conceivable 
that this excess of stars are indeed representatives of the Sgr Stream spread over $\approx$11 kpc ($\Delta m\approx 0.5$) from 42-53\,kpc along 
the line-of-sight. The bottom panel in Fig.~\ref{9kpc_feature} shows the scatter around the isochrone in the combined CMD. 
We attempted to fit the data with a second population but the dominance of the 23\,kpc population and the dispersed
distribution of stars produced only numerically unstable solutions. 
Additional data obtained over a larger field may, however, allow further validation of the Sgr Tidal Stream model and thus its use in interpreting our observations.  In particular, Fig.\ 6 suggests that in the direction of the 1247$-$00 field, the Sgr Stream stars from the distant ($42-53$\,kpc) wrap are relatively frequent.  Observations in this vicinity with wider area coverage may then permit a stronger identification of the predicted Sgr population than is possible with current single field.

\section{Summary and Conclusions}
We report the detection of a prominent population of main sequence stars  in two
Subaru Suprime-Cam fields $1220-01$ and  $1247-00$ located 
in the overlapping region of the Virgo Stellar Stream(VSS) and the Virgo Overdensity (VOD). 
The main sequence stars are at a heliocentric distance of $23.3\pm1.6$\,kpc and have an age of $8.2^{+0.8}_{-0.9}$\,Gyr 
and a mean metallicity of [Fe/H]=$-0.67^{+0.16}_{-0.12}$\,dex as estimates from the best fitting isochrones.
These parameters are similar to the age and metallicity we derived for the main stellar feature in our third 
Suprime-Cam field, $1145+13$, centered on the leading arm of the Sgr Stream, Branch A of the bifurcation. 
The Sgr Stream population has a MSTO color of $(g-r)_0\sim0.25$ and can be traced as faint as  $g\approx25$\,mag in the CMD.
From isochrone fitting we infer an age of $9.1^{+1.0}_{-1.1}$\,Gyr 
and a mean metallicity of [Fe/H]=$-0.70^{+0.15}_{-0.20}$\,dex. 
We believe they are the first measurements of these quantities for the leading tidal tail in this part of the Sgr Stream.
The associated heliocentric distance of $30.9\pm3.0$\,kpc 
agrees well with the estimated distance to the Sgr Tidal Stream in that direction of the Galactic halo 
as inferred from upper main sequence and turn-off stars from SDSS \citep{Belokurov06b}.

The derived parameters for the main sequence stars in the two VSS/VOD fields are also a good 
match to the age of the dominant stellar population (Pop\,A) in the Sagittarius dwarf galaxy \citep[$8.0\pm1.5$\,Gyr;][]{Bellazzini06} 
as well as to the peak in the metallicity distribution function, [Fe/H]=$-0.7$\,dex,  of the $2-3$\,Gyr old M-giants 
in the Sgr north leading arm at the orbital longitude $\Lambda_\sun=260^\circ$ \citep{Chou07}. 
 
The agreement in metallicity may be taken as supporting evidence that the detected VOD main sequence stars are from the Sgr Tidal Stream,
in line with the \cite{MartinezDelgado07} proposition \citep[see also][]{Prior09b} that the Sgr debris is a major 
contributor to the overdensity in Virgo. Although the measured metallicity in the VSS/VOD fields ([Fe/H]=$-0.67^{+0.16}_{-0.12}$\,dex) 
is slightly more metal poor than the stars in the Sgr core where the metallicity distribution function has a spread from $-1.0$ to 
super-solar $+0.2$ and peaks at [Fe/H]=$-0.4$\,dex \citep{Bellazzini08}, the difference is consistent with the abundance 
gradient along the tidal arms reported by \cite{Keller10}. 

However, it must be recalled that metallicity provides only a poor constraint 
as other Milky Way dwarf satellites massive enough to leave a prominent stellar feature in the MW halo can be 
expected to have similar [Fe/H] \citep{KirbyEN08, Meisner12}.  
To test the hypothesis that the VOD is dominated by stars from a different halo structure, 
we compared the CMDs of the VSS/VOD fields with the Sgr Tidal Stream 
model by \cite{Law10} based on a triaxial Galactic halo shape that is empirically calibrated with SDSS Sgr A-branch 
and 2MASS M\,giant stars. In the surveyed directions the model makes precise predictions about the location and strength
of different wraps of the Sgr Tidal Stream: a small number of Sgr Tidal Stream 
stars at a distance of $\approx$9\,kpc and a slightly larger population dispersed over the distance range $42<D<53$\,kpc (Fig.\,\ref{Law10}). Overplotted isochrones on the CMDs at the prescribed distances (Figs.\,\ref{distrange} and \ref{9kpc_feature}) indeed align with features that 
can be interpreted as the predicted Sgr Tidal Stream stars. The predicted number variation of the Sgr Tidal Stream stars between the two fields
was also observed providing additional confidence in the Sgr Tidal Stream model.
A more quantitative analysis of these features by means of isochrone fitting turned out to be unsuccessful because of the 
low contrast. A systematic, deep photometric study over a larger area will be necessary to collect more stars for further insight.
 
The major finding from the analysis of the CMDs of the two selected fields in the complex VSS/VOD region is a prominent main 
sequence population at 23\,kpc. Given these stars are absent in the currently best-fit N-body simulation of the tidal disruption 
of the Sgr dwarf galaxy \citep{Law10} an association with Sagittarius seems unlikely. These stars appear to belong to a different stellar 
substructure in the Galactic halo.
Additionally, we find that at the VOD location this substructure has a similar age and metallicity 
as the Sgr Tidal Stream stars in the leading arm, Branch A of the bifurcation. With a metallicity of [Fe/H]=$-0.67$\,dex the associated 
stars are also significantly more metal-rich than the VSS/VOD RRL stars ([Fe/H]$\approx-1.9$\,dex).

\bigskip 
The authors wish to thank the anonymous referee for providing thoughtful comments that helped 
to improve the paper, and Renee Kraan-Korteweg for reading the early versions of the manuscript. 
HJ and PK acknowledge the financial support from the Go8 - DAAD - Australia/Germany 
Joint Research Co-operation Scheme and HJ and PT acknowledge financial support from the 
Access to Major Research Facilities Programme which is a component of the International 
Science Linkages established under the Australian Government innovation statement, 
Backing Australia's Ability. This research is also supported in part by the Australian Research 
Council through Discovery Projects grants DP0878137 and DP120100475. BW thanks NSF AST 0908446. AJ and 
MZ acknowledge support from BASAL CATA PFB-06, FONDAP CFA 15010003, the Chilean Ministry for the Economy, 
Development, and TourismÕs Programa Iniciativa Cient\'{i}fica Milenio through grant P07-021-F, awarded to The Milky Way Millennium
Nucleus, and Fondecyt REGULAR 1110393. EO acknowledges American NSF support 
through grant AST-0807498. The authors are grateful to David Law and Steven Majewski
for making the details of their model data publicly available.

\end{document}